\documentclass[lettersize,journal]{IEEEtran}
\IEEEoverridecommandlockouts
\usepackage{amsmath,amsfonts}
\usepackage{xcolor}
\usepackage{subcaption}
\usepackage{graphicx}
\usepackage{pgfplots}       
\usepackage{tikz}  
\usepackage{xspace}
\usepackage[acronyms,nonumberlist,nopostdot,nomain,nogroupskip,acronymlists={hidden}]{glossaries}
\usepackage{algorithmic}
\usepackage{algorithm}
\usepackage{array}
\usepackage[caption=false,font=normalsize,labelfont=sf,textfont=sf]{subfig}
\usepackage{textcomp}
\usepackage{stfloats}
\usepackage{url}
\usepackage{verbatim}
\usepackage{graphicx}
\usepackage{cite}
\usepackage{listings}
\usepackage{url}
\usepackage{soul}
\usepackage{balance}

\newglossary[algh]{hidden}{acrh}{acnh}{Hidden Acronyms}

\newcommand{\ric}{\gls{ric}\xspace}

\newcommand{\nearrt}{\gls{nearrt}\xspace}
\newcommand{\nonrt}{\gls{nonrt}\xspace}

\newcommand{\ran}{\gls{ran}\xspace}
\newcommand{\xdev}{xDevSM\xspace}

\definecolor{codegray}{rgb}{0.25,0.25,0.25} 
\definecolor{codepurple}{rgb}{0.58,0,0.82}

\def\mathdefault#1{#1}

\lstdefinestyle{mystile-logs}{
  commentstyle=\color{gray},
  numberstyle=\tiny,
  basicstyle=\ttfamily\scriptsize,
  rulecolor=\color{black},
  breakatwhitespace=true,         
  breaklines=true,                 
  captionpos=b,
  frame=tb,
  keepspaces=true,                 
  numbers=left,                    
  numbersep=5pt,                  
  showspaces=false,                
  showstringspaces=false,
  showtabs=false,                  
  tabsize=2,
  xleftmargin=10pt,
  belowskip=-10pt,
}

\lstdefinelanguage{logs}{
  alsoletter={-},
  keywords={listen,on,off,dst,periodic},
  sensitive=false,
  comment=[l]{\#},
  morecomment=[s]{/*}{*/},
  moredelim=[l][\color{orange}]{\&},
  moredelim=[l][\color{magenta}]{*},
  morestring=[b]',
  morestring=[b]",
}

\lstdefinelanguage{pythonlite}{
  keywords={def,class,if,else,elif,for,while,return,import,from,as,with,try,except,finally,True,False,None},
  alsoletter={-},
  sensitive=true,
  comment=[l]{\#},
  morestring=[b]',
  morestring=[b]",
}

\lstdefinestyle{mypython}{
  language=pythonlite,
  numberstyle=\tiny,
  basicstyle=\ttfamily\scriptsize,
  keywordstyle=\color{blue},
  commentstyle=\color{gray},
  rulecolor=\color{black},
  breakatwhitespace=true,         
  breaklines=true,                 
  captionpos=b,
  frame=tb,
  keepspaces=true,                 
  numbers=left,                    
  numbersep=5pt,                  
  showspaces=false,                
  showstringspaces=false,
  showtabs=false,                  
  tabsize=2,
  xleftmargin=10pt,
  belowskip=-10pt,
}

\pgfplotsset{compat=1.18}

\newif\ifexttikz
\exttikzfalse
\ifexttikz
\else
\usepackage{tikzpagenodes,etoolbox}
\usetikzlibrary{calc}
\usepackage[contents={}]{background}
\AddEverypageHook{%
\ifnumequal{\thepage}{1}{%
    \tikz[remember picture,overlay]{%
        \node[draw,
        minimum width=1.03\textwidth,
        text width=1.02\textwidth,
        font=\footnotesize
        ]
        at ($(current page header area) - (0,5pt)$)
        {%
        This paper has been submitted on IEEE Transactions on Network and Service Management for possible publication.  Copyright may be transferred without notice, after which this version may
no longer be accessible.
        };
    }%
}{}
}
\fi

\begin{document}


\title{xDevSM: An Open-Source Framework for Portable, AI-Ready xApps Across Heterogeneous O-RAN Deployments}

\author{Angelo Feraudo*,
Stefano Maxenti,
Andrea Lacava,
Leonardo Bonati,\\
Paolo Bellavista,
Michele Polese,
Tommaso Melodia%
\thanks{Manuscript received April 19, 2021; revised August 16, 2021.}%
\thanks{* indicates the corresponding author.}%
\thanks{Angelo Feraudo and Paolo Bellavista are with the Department of Computer Science and Engineering, University of Bologna, Italy (e-mail: \{first.last\}@unibo.it).\\
Angelo Feraudo, Stefano Maxenti, Andrea Lacava, Leonardo Bonati, Michele Polese, and Tommaso Melodia are with the Institute for the Wireless Internet of Things, Northeastern University, Boston, MA, U.S.A. (e-mail: maxenti.s@northeastern.edu; \{f.last\}@northeastern.edu).}
\thanks{
This article is based upon work partially supported by the National Telecommunications and Information Administration (NTIA)'s Public Wireless Supply Chain Innovation Fund (PWSCIF) under Award No. 25-60-IF054, by the U.S. NSF under grant TI-2449452, and by the European Union’s NextGenerationEU instrument, under the Italian National Recovery and Resilience Plan (NRRP), Mission 4 Component 2 Investment 1.3, enlarged partnership “Telecommunications of the Future” (PE0000001), program “RESTART”.
} 
}

\markboth{Journal of \LaTeX\ Class Files,~Vol.~14, No.~8, August~2021}%
{Shell \MakeLowercase{\textit{et al.}}: A Sample Article Using IEEEtran.cls for IEEE Journals}

\newacronym{3gpp}{3GPP}{3rd Generation Partnership Project}
\newacronym{4g}{4G}{4th generation}
\newacronym{5g}{5G}{5th generation}
\newacronym{6g}{6G}{6th generation}
\newacronym{5gc}{5GC}{5G Core}
\newacronym{adc}{ADC}{Analog to Digital Converter}
\newacronym{aerpaw}{AERPAW}{Aerial Experimentation and Research Platform for Advanced Wireless}
\newacronym{ai}{AI}{Artificial Intelligence}
\newacronym{aimd}{AIMD}{Additive Increase Multiplicative Decrease}
\newacronym{am}{AM}{Acknowledged Mode}
\newacronym{amc}{AMC}{Adaptive Modulation and Coding}
\newacronym{amf}{AMF}{Access and Mobility Management Function}
\newacronym{aops}{AOPS}{Adaptive Order Prediction Scheduling}
\newacronym{api}{API}{Application Programming Interface}
\newacronym{apn}{APN}{Access Point Name}
\newacronym{ap}{AP}{Application Protocol}
\newacronym{aqm}{AQM}{Active Queue Management}
\newacronym{ausf}{AUSF}{Authentication Server Function}
\newacronym{avc}{AVC}{Advanced Video Coding}
\newacronym{awgn}{AGWN}{Additive White Gaussian Noise}
\newacronym{balia}{BALIA}{Balanced Link Adaptation Algorithm}
\newacronym{bbu}{BBU}{Base Band Unit}
\newacronym{bdp}{BDP}{Bandwidth-Delay Product}
\newacronym{ber}{BER}{Bit Error Rate}
\newacronym{bf}{BF}{Beamforming}
\newacronym{bler}{BLER}{Block Error Rate}
\newacronym{brr}{BRR}{Bayesian Ridge Regressor}
\newacronym{bs}{BS}{Base Station}
\newacronym{bsr}{BSR}{Buffer Status Report}
\newacronym{bss}{BSS}{Business Support System}
\newacronym{ca}{CA}{Carrier Aggregation}
\newacronym{caas}{CaaS}{Connectivity-as-a-Service}
\newacronym{cb}{CB}{Code Block}
\newacronym{cc}{CC}{Congestion Control}
\newacronym{ccid}{CCID}{Congestion Control ID}
\newacronym{cco}{CC}{Carrier Component}
\newacronym{ci}{CI}{Continuous Integration}
\newacronym{cd}{CD}{Continuous Delivery}
\newacronym{cdd}{CDD}{Cyclic Delay Diversity}
\newacronym{cdf}{CDF}{Cumulative Distribution Function}
\newacronym{cdn}{CDN}{Content Distribution Network}
\newacronym{cli}{CLI}{Command-line Interface}
\newacronym{cn}{CN}{Core Network}
\newacronym{codel}{CoDel}{Controlled Delay Management}
\newacronym{comac}{COMAC}{Converged Multi-Access and Core}
\newacronym{cord}{CORD}{Central Office Re-architected as a Datacenter}
\newacronym{cornet}{CORNET}{COgnitive Radio NETwork}
\newacronym{cosmos}{COSMOS}{Cloud Enhanced Open Software Defined Mobile Wireless Testbed for City-Scale Deployment}
\newacronym{cots}{COTS}{Commercial Off-the-Shelf}
\newacronym{cp}{CP}{Control Plane}
\newacronym{cyp}{CP}{Cyclic Prefix}
\newacronym{up}{UP}{User Plane}
\newacronym{cpu}{CPU}{Central Processing Unit}
\newacronym{cqi}{CQI}{Channel Quality Information}
\newacronym{cr}{CR}{Cognitive Radio}
\newacronym{cran}{CRAN}{Cloud \gls{ran}}
\newacronym{crs}{CRS}{Cell Reference Signal}
\newacronym{csi}{CSI}{Channel State Information}
\newacronym{csirs}{CSI-RS}{Channel State Information - Reference Signal}
\newacronym{cu}{CU}{Central Unit}
\newacronym{d2tcp}{D$^2$TCP}{Deadline-aware Data center TCP}
\newacronym{d3}{D$^3$}{Deadline-Driven Delivery}
\newacronym{dac}{DAC}{Digital to Analog Converter}
\newacronym{dag}{DAG}{Directed Acyclic Graph}
\newacronym{das}{DAS}{Distributed Antenna System}
\newacronym{dash}{DASH}{Dynamic Adaptive Streaming over HTTP}
\newacronym{dc}{DC}{Dual Connectivity}
\newacronym{dccp}{DCCP}{Datagram Congestion Control Protocol}
\newacronym{dce}{DCE}{Direct Code Execution}
\newacronym{dci}{DCI}{Downlink Control Information}
\newacronym{dctcp}{DCTCP}{Data Center TCP}
\newacronym{dl}{DL}{Downlink}
\newacronym{dmr}{DMR}{Deadline Miss Ratio}
\newacronym{dmrs}{DMRS}{DeModulation Reference Signal}
\newacronym{drb}{DRB}{Data Radio Bearer}
\newacronym{drlcc}{DRL-CC}{Deep Reinforcement Learning Congestion Control}
\newacronym{drs}{DRS}{Discovery Reference Signal}
\newacronym{du}{DU}{Distributed Unit}
\newacronym{e2e}{E2E}{end-to-end}
\newacronym{earfcn}{EARFCN}{E-UTRA Absolute Radio Frequency Channel Number}
\newacronym{ecaas}{ECaaS}{Edge-Cloud-as-a-Service}
\newacronym{ecn}{ECN}{Explicit Congestion Notification}
\newacronym{edf}{EDF}{Earliest Deadline First}
\newacronym{embb}{eMBB}{Enhanced Mobile Broadband}
\newacronym{empower}{EMPOWER}{EMpowering transatlantic PlatfOrms for advanced WirEless Research}
\newacronym{enb}{eNB}{evolved Node Base}
\newacronym{endc}{EN-DC}{E-UTRAN-\gls{nr} \gls{dc}}
\newacronym{epc}{EPC}{Evolved Packet Core}
\newacronym{eps}{EPS}{Evolved Packet System}
\newacronym{es}{ES}{Edge Server}
\newacronym{etsi}{ETSI}{European Telecommunications Standards Institute}
\newacronym[firstplural=Estimated Times of Arrival (ETAs)]{eta}{ETA}{Estimated Time of Arrival}
\newacronym{eutran}{E-UTRAN}{Evolved Universal Terrestrial Access Network}
\newacronym{faas}{FaaS}{Function-as-a-Service}
\newacronym{fapi}{FAPI}{Functional Application Platform Interface}
\newacronym{fdd}{FDD}{Frequency Division Duplexing}
\newacronym{fdm}{FDM}{Frequency Division Multiplexing}
\newacronym{fdma}{FDMA}{Frequency Division Multiple Access}
\newacronym{fed4fire}{FED4FIRE+}{Federation 4 Future Internet Research and Experimentation Plus}
\newacronym{fir}{FIR}{Finite Impulse Response}
\newacronym{fit}{FIT}{Future \acrlong{iot}}
\newacronym{fpga}{FPGA}{Field Programmable Gate Array}
\newacronym{fr2}{FR2}{Frequency Range 2}
\newacronym{fs}{FS}{Fast Switching}
\newacronym{fscc}{FSCC}{Flow Sharing Congestion Control}
\newacronym{ftp}{FTP}{File Transfer Protocol}
\newacronym{fw}{FW}{Flow Window}
\newacronym{ge}{GE}{Gaussian Elimination}
\newacronym{gnb}{gNB}{Next Generation Node Base}
\newacronym{gop}{GOP}{Group of Pictures}
\newacronym{gpr}{GPR}{Gaussian Process Regressor}
\newacronym{gpu}{GPU}{Graphics Processing Unit}
\newacronym{gtp}{GTP}{GPRS Tunneling Protocol}
\newacronym{gtpc}{GTP-C}{GPRS Tunnelling Protocol Control Plane}
\newacronym{gtpu}{GTP-U}{GPRS Tunnelling Protocol User Plane}
\newacronym{gtpv2c}{GTPv2-C}{\gls{gtp} v2 - Control}
\newacronym{gw}{GW}{Gateway}
\newacronym{harq}{HARQ}{Hybrid Automatic Repeat reQuest}
\newacronym{hetnet}{HetNet}{Heterogeneous Network}
\newacronym{hh}{HH}{Hard Handover}
\newacronym{hol}{HOL}{Head-of-Line}
\newacronym{hqf}{HQF}{Highest-quality-first}
\newacronym{hss}{HSS}{Home Subscription Server}
\newacronym{http}{HTTP}{HyperText Transfer Protocol}
\newacronym{ia}{IA}{Initial Access}
\newacronym{iab}{IAB}{Integrated Access and Backhaul}
\newacronym{ic}{IC}{Incident Command}
\newacronym{ietf}{IETF}{Internet Engineering Task Force}
\newacronym{imsi}{IMSI}{International Mobile Subscriber Identity}
\newacronym{imt}{IMT}{International Mobile Telecommunication}
\newacronym{iot}{IoT}{Internet of Things}
\newacronym{ip}{IP}{Internet Protocol}
\newacronym{itu}{ITU}{International Telecommunication Union}
\newacronym{kpi}{KPI}{Key Performance Indicator}
\newacronym{kpm}{KPM}{Key Performance Measurement}
\newacronym{kvm}{KVM}{Kernel-based Virtual Machine}
\newacronym{los}{LOS}{Line-of-Sight}
\newacronym{lsm}{LSM}{Link-to-System Mapping}
\newacronym{lstm}{LSTM}{Long Short Term Memory}
\newacronym{lte}{LTE}{Long Term Evolution}
\newacronym{lxc}{LXC}{Linux Container}
\newacronym{m2m}{M2M}{Machine to Machine}
\newacronym{mac}{MAC}{Medium Access Control}
\newacronym{manet}{MANET}{Mobile Ad Hoc Network}
\newacronym{mano}{MANO}{Management and Orchestration}
\newacronym{mc}{MC}{Multi-Connectivity}
\newacronym{mcc}{MCC}{Mobile Cloud Computing}
\newacronym{mchem}{MCHEM}{Massive Channel Emulator}
\newacronym{mcs}{MCS}{Modulation and Coding Scheme}
\newacronym{mec}{MEC}{Multi-access Edge Computing}
\newacronym{mec2}{MEC}{Mobile Edge Cloud}
\newacronym{mfc}{MFC}{Mobile Fog Computing}
\newacronym{mgen}{MGEN}{Multi-Generator}
\newacronym{mi}{MI}{Mutual Information}
\newacronym{mib}{MIB}{Master Information Block}
\newacronym{miesm}{MIESM}{Mutual Information Based Effective SINR}
\newacronym{mimo}{MIMO}{Multiple Input, Multiple Output}
\newacronym{ml}{ML}{Machine Learning}
\newacronym{mlr}{MLR}{Maximum-local-rate}
\newacronym[plural=\gls{mme}s,firstplural=Mobility Management Entities (MMEs)]{mme}{MME}{Mobility Management Entity}
\newacronym{mmtc}{mMTC}{Massive Machine-Type Communications}
\newacronym{mmwave}{mmWave}{millimeter wave}
\newacronym{mpdccp}{MP-DCCP}{Multipath Datagram Congestion Control Protocol}
\newacronym{mptcp}{MPTCP}{Multipath TCP}
\newacronym{mr}{MR}{Maximum Rate}
\newacronym{mrdc}{MR-DC}{Multi \gls{rat} \gls{dc}}
\newacronym{mse}{MSE}{Mean Square Error}
\newacronym{mss}{MSS}{Maximum Segment Size}
\newacronym{mt}{MT}{Mobile Termination}
\newacronym{mtd}{MTD}{Machine-Type Device}
\newacronym{mtu}{MTU}{Maximum Transmission Unit}
\newacronym{mumimo}{MU-MIMO}{Multi-user \gls{mimo}}
\newacronym{mvno}{MVNO}{Mobile Virtual Network Operator}
\newacronym{nalu}{NALU}{Network Abstraction Layer Unit}
\newacronym{nas}{NAS}{Network Attached Storage}
\newacronym{nat}{NAT}{Network Address Translation}
\newacronym{nbiot}{NB-IoT}{Narrow Band IoT}
\newacronym{nfv}{NFV}{Network Function Virtualization}
\newacronym{nfvi}{NFVI}{Network Function Virtualization Infrastructure}
\newacronym{ni}{NI}{Network Interfaces}
\newacronym{nic}{NIC}{Network Interface Card}
\newacronym{nlos}{NLOS}{Non-Line-of-Sight}
\newacronym{now}{NOW}{Non Overlapping Window}
\newacronym{nsm}{NSM}{Network Service Mesh}
\newacronym[type=hidden]{nr}{NR}{New Radio}
\newacronym{nrf}{NRF}{Network Repository Function}
\newacronym{nsa}{NSA}{Non Stand Alone}
\newacronym{nse}{NSE}{Network Slicing Engine}
\newacronym{nssf}{NSSF}{Network Slice Selection Function}
\newacronym{o2i}{O2I}{Outdoor to Indoor}
\newacronym{oai}{OAI}{OpenAirInterface}
\newacronym{oaicn}{OAI-CN}{\gls{oai} \acrlong{cn}}
\newacronym{oairan}{OAI-RAN}{\acrlong{oai} \acrlong{ran}}
\newacronym{oam}{OAM}{Operations, Administration and Maintenance}
\newacronym{ofdm}{OFDM}{Orthogonal Frequency Division Multiplexing}
\newacronym{olia}{OLIA}{Opportunistic Linked Increase Algorithm}
\newacronym{omec}{OMEC}{Open Mobile Evolved Core}
\newacronym{onap}{ONAP}{Open Network Automation Platform}
\newacronym{onf}{ONF}{Open Networking Foundation}
\newacronym{onos}{ONOS}{Open Networking Operating System}
\newacronym{oom}{OOM}{\gls{onap} Operations Manager}
\newacronym{opnfv}{OPNFV}{Open Platform for \gls{nfv}}
\newacronym[type=hidden]{oran}{O-RAN}{Open RAN}
\newacronym{orbit}{ORBIT}{Open-Access Research Testbed for Next-Generation Wireless Networks}
\newacronym{os}{OS}{Operating System}
\newacronym{oss}{OSS}{Operations Support System}
\newacronym{pa}{PA}{Position-aware}
\newacronym{pase}{PASE}{Prioritization, Arbitration, and Self-adjusting Endpoints}
\newacronym{pawr}{PAWR}{Platforms for Advanced Wireless Research}
\newacronym{pbch}{PBCH}{Physical Broadcast Channel}
\newacronym{pcef}{PCEF}{Policy and Charging Enforcement Function}
\newacronym{pcfich}{PCFICH}{Physical Control Format Indicator Channel}
\newacronym{pcrf}{PCRF}{Policy and Charging Rules Function}
\newacronym{pdcch}{PDCCH}{Physical Downlink Control Channel}
\newacronym{pdcp}{PDCP}{Packet Data Convergence Protocol}
\newacronym{pdf}{PDF}{Probability Density Function}
\newacronym{pdsch}{PDSCH}{Physical Downlink Shared Channel}
\newacronym{pdu}{PDU}{Packet Data Unit}
\newacronym{pf}{PF}{Proportional Fair}
\newacronym{pgw}{PGW}{Packet Gateway}
\newacronym{phich}{PHICH}{Physical Hybrid ARQ Indicator Channel}
\newacronym{phy}{PHY}{Physical}
\newacronym{pmch}{PMCH}{Physical Multicast Channel}
\newacronym{pmi}{PMI}{Precoding Matrix Indicators}
\newacronym{powder}{POWDER}{Platform for Open Wireless Data-driven Experimental Research}
\newacronym{ppo}{PPO}{Proximal Policy Optimization}
\newacronym{ppp}{PPP}{Poisson Point Process}
\newacronym{prach}{PRACH}{Physical Random Access Channel}
\newacronym{prb}{PRB}{Physical Resource Block}
\newacronym{psnr}{PSNR}{Peak Signal to Noise Ratio}
\newacronym{pss}{PSS}{Primary Synchronization Signal}
\newacronym{pucch}{PUCCH}{Physical Uplink Control Channel}
\newacronym{pusch}{PUSCH}{Physical Uplink Shared Channel}
\newacronym{qam}{QAM}{Quadrature Amplitude Modulation}
\newacronym{qci}{QCI}{\gls{qos} Class Identifier}
\newacronym{qoe}{QoE}{Quality of Experience}
\newacronym{qos}{QoS}{Quality of Service}
\newacronym{quic}{QUIC}{Quick UDP Internet Connections}
\newacronym{rach}{RACH}{Random Access Channel}
\newacronym{ran}{RAN}{Radio Access Network}

\newacronym[firstplural=Radio Access Technologies (RATs)]{rat}{RAT}{Radio Access Technology}
\newacronym{rbg}{RBG}{Resource Block Group}
\newacronym{rcn}{RCN}{Research Coordination Network}
\newacronym{rc}{RC}{RAN Control}
\newacronym{rec}{REC}{Radio Edge Cloud}
\newacronym{red}{RED}{Random Early Detection}
\newacronym{renew}{RENEW}{Reconfigurable Eco-system for Next-generation End-to-end Wireless}
\newacronym{rf}{RF}{Radio Frequency}
\newacronym{rfc}{RFC}{Request for Comments}
\newacronym{rfr}{RFR}{Random Forest Regressor}
\newacronym{ric}{RIC}{RAN Intelligent Controller}
\newacronym{nrric}{Near-RT RIC}{Near-Real-Time RAN Intelligent Controller}
\newacronym{rl}{RL}{Reinforcement Learning}
\newacronym{rlc}{RLC}{Radio Link Control}
\newacronym{rlf}{RLF}{Radio Link Failure}
\newacronym{rlnc}{RLNC}{Random Linear Network Coding}
\newacronym{rmr}{RMR}{RIC Message Router}
\newacronym{rmse}{RMSE}{Root Mean Squared Error}
\newacronym{rnis}{RNIS}{Radio Network Information Service}
\newacronym{rr}{RR}{Round Robin}
\newacronym{rrc}{RRC}{Radio Resource Control}
\newacronym{rrm}{RRM}{Radio Resource Management}
\newacronym{rru}{RRU}{Remote Radio Unit}
\newacronym{rs}{RS}{Remote Server}
\newacronym{rsrp}{RSRP}{Reference Signal Received Power}
\newacronym{rsrq}{RSRQ}{Reference Signal Received Quality}
\newacronym{rss}{RSS}{Received Signal Strength}
\newacronym{rssi}{RSSI}{Received Signal Strength Indicator}
\newacronym{rtt}{RTT}{Round Trip Time}
\newacronym{ru}{RU}{Radio Unit}
\newacronym{rw}{RW}{Receive Window}
\newacronym{rx}{RX}{Receiver}
\newacronym{snssai}{S-NSSAI}{Single Network Slice Selection Assistance Information}
\newacronym{s1ap}{S1AP}{S1 Application Protocol}
\newacronym{sa}{SA}{standalone}
\newacronym{sack}{SACK}{Selective Acknowledgment}
\newacronym{sap}{SAP}{Service Access Point}
\newacronym{sc2}{SC2}{Spectrum Collaboration Challenge}
\newacronym{scef}{SCEF}{Service Capability Exposure Function}
\newacronym{sch}{SCH}{Secondary Cell Handover}
\newacronym{scoot}{SCOOT}{Split Cycle Offset Optimization Technique}
\newacronym{sctp}{SCTP}{Stream Control Transmission Protocol}
\newacronym{sdap}{SDAP}{Service Data Adaptation Protocol}
\newacronym{sdk}{SDK}{Software Development Kit}
\newacronym{sdm}{SDM}{Space Division Multiplexing}
\newacronym{sdma}{SDMA}{Spatial Division Multiple Access}
\newacronym{sdl}{SDL}{Shared Data Layer}
\newacronym{sdn}{SDN}{Software-defined Networking}
\newacronym{sdr}{SDR}{Software-defined Radio}
\newacronym{seba}{SEBA}{SDN-Enabled Broadband Access}
\newacronym{sgsn}{SGSN}{Serving GPRS Support Node}
\newacronym{sgw}{SGW}{Service Gateway}
\newacronym{si}{SI}{Study Item}
\newacronym{sib}{SIB}{Secondary Information Block}
\newacronym{sinr}{SINR}{Signal to Interference plus Noise Ratio}
\newacronym{sip}{SIP}{Session Initiation Protocol}
\newacronym{siso}{SISO}{Single Input, Single Output}
\newacronym{sla}{SLA}{Service Level Agreement}
\newacronym{sm}{SM}{Service Model}
\newacronym{e2sm}{E2SM}{E2 Service Model}
\newacronym{e2ap}{E2AP}{E2 Application Protocol}
\newacronym{smf}{SMF}{Session Management Function}
\newacronym{smo}{SMO}{Service Management and Orchestration}
\newacronym{sms}{SMS}{Short Message Service}
\newacronym{smsgmsc}{SMS-GMSC}{\gls{sms}-Gateway}
\newacronym{snr}{SNR}{Signal-to-Noise-Ratio}
\newacronym{son}{SON}{Self-Organizing Network}
\newacronym{sptcp}{SPTCP}{Single Path TCP}
\newacronym{srb}{SRB}{Service Radio Bearer}
\newacronym{srn}{SRN}{Standard Radio Node}
\newacronym{srs}{SRS}{Sounding Reference Signal}
\newacronym{ss}{SS}{Synchronization Signal}
\newacronym{sss}{SSS}{Secondary Synchronization Signal}
\newacronym{st}{ST}{Spanning Tree}
\newacronym{svc}{SVC}{Scalable Video Coding}
\newacronym{tb}{TB}{Transport Block}
\newacronym{tcp}{TCP}{Transmission Control Protocol}
\newacronym{tdd}{TDD}{Time Division Duplexing}
\newacronym{tdm}{TDM}{Time Division Multiplexing}
\newacronym{tdma}{TDMA}{Time Division Multiple Access}
\newacronym{tfl}{TfL}{Transport for London}
\newacronym{tfrc}{TFRC}{TCP-Friendly Rate Control}
\newacronym{tft}{TFT}{Traffic Flow Template}
\newacronym{tgen}{TGEN}{Traffic Generator}
\newacronym{tip}{TIP}{Telecom Infra Project}
\newacronym{tm}{TM}{Transparent Mode}
\newacronym{to}{TO}{Telco Operator}
\newacronym{tr}{TR}{Technical Report}
\newacronym{trp}{TRP}{Transmitter Receiver Pair}
\newacronym{ts}{TS}{Technical Specification}
\newacronym{tti}{TTI}{Transmission Time Interval}
\newacronym{ttt}{TTT}{Time-to-Trigger}
\newacronym{tx}{TX}{Transmitter}
\newacronym{uas}{UAS}{Unmanned Aerial System}
\newacronym{uav}{UAV}{Unmanned Aerial Vehicle}
\newacronym{udm}{UDM}{Unified Data Management}
\newacronym{udp}{UDP}{User Datagram Protocol}
\newacronym{udr}{UDR}{Unified Data Repository}
\newacronym{ue}{UE}{User Equipment}
\newacronym{uhd}{UHD}{\gls{usrp} Hardware Driver}
\newacronym{ul}{UL}{Uplink}
\newacronym{um}{UM}{Unacknowledged Mode}
\newacronym{uml}{UML}{Unified Modeling Language}
\newacronym{upa}{UPA}{Uniform Planar Array}
\newacronym{upf}{UPF}{User Plane Function}
\newacronym{urllc}{URLLC}{Ultra Reliable and Low Latency Communications}
\newacronym{usa}{U.S.}{United States}
\newacronym{usim}{USIM}{Universal Subscriber Identity Module}
\newacronym{usrp}{USRP}{Universal Software Radio Peripheral}
\newacronym{utc}{UTC}{Urban Traffic Control}
\newacronym{vim}{VIM}{Virtualization Infrastructure Manager}
\newacronym{vm}{VM}{Virtual Machine}
\newacronym{vnf}{VNF}{Virtual Network Function}
\newacronym{volte}{VoLTE}{Voice over \gls{lte}}
\newacronym{voltha}{VOLTHA}{Virtual OLT HArdware Abstraction}
\newacronym{vr}{VR}{Virtual Reality}
\newacronym{vran}{vRAN}{Virtualized \gls{ran}}
\newacronym{vss}{VSS}{Video Streaming Server}
\newacronym{wbf}{WBF}{Wired Bias Function}
\newacronym{wf}{WF}{Waterfilling}
\newacronym{wg}{WG}{Working Group}
\newacronym{wlan}{WLAN}{Wireless Local Area Network}
\newacronym{osm}{OSM}{Open Source \gls{nfv} Management and Orchestration}
\newacronym{pnf}{PNF}{Physical Network Function}
\newacronym{drl}{DRL}{Deep Reinforcement Learning}
\newacronym{mtc}{MTC}{Machine-type Communications}
\newacronym{osc}{OSC}{O-RAN Software Community}
\newacronym{mns}{MnS}{Management Services}
\newacronym{ves}{VES}{\gls{vnf} Event Stream}
\newacronym{ei}{EI}{Enrichment Information}
\newacronym{fh}{FH}{Fronthaul}
\newacronym{fft}{FFT}{Fast Fourier Transform}
\newacronym{laa}{LAA}{Licensed-Assisted Access}
\newacronym{plfs}{PLFS}{Physical Layer Frequency Signals}
\newacronym{ptp}{PTP}{Precision Time Protocol}
\newacronym{cbrs}{CBRS}{Citizen Broadband Radio Service}
\newacronym{rnti}{RNTI}{Radio Network Temporary Identifier}
\newacronym{tbs}{TBS}{Transport Block Size}

\newacronym{onr}{ONR}{Office of Naval Research}
\newacronym{afosr}{AFOSR}{Air Force Office of Scientific Research}
\newacronym{afrl}{AFRL}{Air Force Research Laboratory}
\newacronym{arl}{ARL}{Army Research Laboratory}

\newacronym{ct}{CT}{Continuous Testing}
\newacronym{mno}{MNO}{Mobile Network Operator}
\newacronym{oci}{OCI}{Open Container Initiative}
\newacronym{macsec}{MACsec}{Media Access Control Security}
\newacronym{pt}{PT}{Plain Text}
\newacronym{cuda}{CUDA}{Compute Unified Device Architecture}
\newacronym{dsp}{DSP}{Digital Signal Processing}

\newacronym{cus}{CUS}{Control, User, Synchronization}
\newacronym{dpd}{DPD}{Digital Pre-Distorsion}
\newacronym{cfr}{CFR}{Crest Factor Reduction}
\newacronym{pci}{PCIe}{Peripheral Component Interconnect Express}
\newacronym{dpu}{DPU}{Data Processing Unit}
\newacronym{rfsoc}{RFSoC}{Radio Frequency System-on-Chip}
\newacronym{if}{IF}{Intermediate Frequency}
\newacronym{nyu}{NYU}{New York University}
\newacronym{gh}{GH}{Grace Hopper}
\newacronym{trl}{TRL}{Technology Readiness Level}
\newacronym{srfa}{SRFA}{Special Research Focus Area}
\newacronym{qsfp}{QSFP}{quad small form factor pluggable}
\newacronym{pse}{PSE}{Performance Specialized Engine}
\newacronym{cae}{CAE}{Cognitive Analysis Engine}
\newacronym{simd}{SIMD}{Single Instruction/Multiple Data}
\newacronym{rt}{RT}{Real-Time}
\newacronym{asm}{ASM}{Advanced Sleep Mode}
\newacronym{aoa}{AoA}{Angle of Arrival}
\newacronym{eaxcid}{eAxC\_ID}{extended Antenna-Carrier Identifier}
\newacronym{xr}{XR}{Extended Reality}
\newacronym{bwp}{BWP}{Bandwidth Part}
\newacronym{dfe}{DFE}{Digital Front-End}
\newacronym{spi}{SPI}{Serial Peripheral Interface}
\newacronym{gpio}{GPIO}{General Purpose Input/Output}
\newacronym{nco}{NCO}{Numerically Controlled Oscillator}
\newacronym{lo}{LO}{Local Oscillator}
\newacronym{lna}{LNA}{Low-Noise Amplifier}
\newacronym{pll}{PLL}{Phased-Locked Loop}
\newacronym{som}{SOM}{System-on-Module}
\newacronym{papr}{PAPR}{Peak-to-Average Power Ratio}
\newacronym{pcb}{PCB}{Printed Circuit Board}
\newacronym{gcpw}{GCPW}{Grounded Co-Planar Waveguide}
\newacronym{cnn}{CNN}{Convolutional Neural Network}
\newacronym{gmp}{GMP}{Generalized Memory Polynomial}
\newacronym{ngrg}{nGRG}{next Generation Research Group}
\newacronym{mrl}{MRL}{Manufacturing Readiness Level}
\newacronym{fr}{FR}{Frequency Range}
\newacronym{sst}{SST}{Slice/Service Type}
\newacronym{sd}{SD}{Service Differentiator}

\newacronym{sbom}{SBOM}{Software Bill of Materials}
\newacronym{hbom}{HBOM}{Hardware Bill of Materials}
\newacronym{vex}{VEX}{Vulnerability Exploitability eXchange}
\newacronym{dos}{DoS}{Denial of Service}
\newacronym{sme}{SME}{Small-Medium Enterprise}

\newacronym{ulpi}{ULPI}{Uplink Performance Improvement}
\newacronym{oem}{OEM}{Original Equipment Manufacturer}
\newacronym{nsin}{NSIN}{National Security Innovation Network}
\newacronym{dod}{DoD}{Department of Defense}
\newacronym{arpu}{ARPU}{Average Revenue per User}
\newacronym{opex}{OPEX}{operational expenses}
\newacronym{txb}{TXB}{Transmit Beam}
\newacronym{cve}{CVE}{Common Vulnerabilities and Exposure}
\newacronym{arc}{ARC}{Aerial RAN CoLab}

\newacronym{ota}{OTA}{Over-the-Air}
\newacronym{nearrt}{Near-RT}{Near-Real-Time}
\newacronym{nonrt}{Non-RT}{Non-Real-Time}

\newacronym{rbc}{RBC}{Radio Bearer Control}
\newacronym{rrac}{RRAC}{Radio Resource Allocation Control}
\newacronym{cmmc}{CMMC}{Connected Mode Mobility Control}
\newacronym{ccc}{CCC}{Cell Configuration and Control}
\newacronym{llc}{LLC}{Lower Layer Control}

\maketitle

\begin{abstract}

Openness and programmability in the O-RAN architecture enable closed-loop control of the \gls{ran}. 
\gls{ai}-driven xApps, in the near-real-time \gls{ric}, 
can learn from network data, anticipate future conditions, and dynamically adapt radio configurations. However, their development and adoption are hindered by the complexity of low-level \gls{ran} control and monitoring message models exposed over the O-RAN E2 interface, limited interoperability across heterogeneous \gls{ran} software stacks, and the lack of developer-friendly frameworks.
In this paper, we introduce \xdev, a framework that significantly lowers the barrier to xApp development by unifying observability and control in O-RAN deployment. By exposing a rich set of \glspl{kpm} and enabling fine-grained radio resource management controls, \xdev provides the essential foundation for practical \gls{ai}-driven xApps. We validate \xdev on real-world testbeds, leveraging \gls{cots} devices together with heterogeneous \gls{ran} hardware, including \gls{usrp}-based \glspl{sdr} and Foxconn radio units, and show its seamless interoperability across multiple open-source \gls{ran} software stacks. Furthermore, we discuss and evaluate the capabilities of our framework through three O-RAN-based scenarios of high interest: (i) \gls{kpm}-based monitoring of network performance, (ii) slice-level \gls{prb} allocation control across multiple \glspl{ue} and slices, and (iii) mobility-aware handover control, showing that \xdev can implement intelligent closed-loop applications, laying the groundwork for learning-based optimization in heterogeneous \gls{ran} deployments. \xdev is open source and available as foundational tool for the research community. 
\end{abstract}


\glsresetall

\markboth{}{}

\section{Introduction}
\label{sec:intro}
\IEEEPARstart{T}{he} rapid proliferation of data-centric and automated services in \glspl{ran} is driving network infrastructures toward unprecedented levels of complexity, challenging the management and control capabilities of current \gls{5g} deployments~\cite{chowdhury20206g,giordani2020toward}. While \gls{5g} introduces enhanced capacity, flexibility, and support for diverse service requirements, emerging use cases already demand more adaptive, intelligent, and automated control mechanisms. Looking ahead, \gls{6g} is envisioned not merely as an evolution of \gls{5g}, but as an enabler of future digital ecosystems, supporting immersive applications such as \gls{xr}, tactile internet, and pervasive connectivity, while offering native integration of \gls{ai} and intelligence-driven automation.

Central to this evolution are the softwarization and virtualization of network functions, which provide the flexibility needed to adapt to highly dynamic environments required in both advanced \gls{5g} and future \gls{6g} systems. To this end, the O-RAN architecture has emerged as a key enabler of open and programmable networks~\cite{polese2022understanding}, with the O-RAN ALLIANCE driving specifications that place openness and intelligence at the core of next-generation \gls{ran} design. By leveraging these principles, O-RAN introduces disaggregated and virtualized components that interact via open and standardized interfaces. 
It defines two \glspl{ric}---the \nonrt and \nearrt \gls{ric}---that operate at different time scales and act upon various \gls{ran} parameters. The \nonrt \gls{ric} supports control loops with time scales larger than $1$\:s, and hosts rApps that influence the \gls{ran} through higher-level policies. These are communicated to the \nearrt \gls{ric} via the A1 interface, which also supports the \gls{ai}/\gls{ml} model life cycle.
%
The \nearrt \gls{ric} operates on time scales ranging from $10$:ms to $1$:s. It hosts xApps that perform fine-grained, near–real-time control of the \gls{ran}. The \gls{ric} communicates directly with the \gls{ran} through the E2 interface. This interface uses the \gls{e2ap} for signaling and control procedures. Additionally, it provides a set of service models that define the measurements and control actions available to xApps.
Unlike rApps, which are limited to policy definition and high-level analytics at the \nonrt \gls{ric}, xApps run on the \nearrt \gls{ric} and interface directly with the \gls{ran} to perform detailed radio resource management and control.
In particular, \gls{ai}-driven xApps have emerged as a promising approach to enhance closed-loop control by leveraging network data to identify patterns, predict future conditions, and dynamically adapt radio configurations. For instance, recent studies have demonstrated that such intelligence-driven applications can significantly improve spectrum utilization, throughput, and user quality of experience~\cite{polese2022colo,puligheddu2023semoran,zangooei2024flexible,irazabal2024tc,doro2024orchestran}.


These advances have stimulated the development of several open-source projects that target the \nearrt \gls{ric} and its integration with disaggregated \glspl{gnb} through the E2 interface~\cite{santos2025managing}. For instance, \gls{osc} provides an open-source implementation of a \nearrt \ric~\cite{nearrtric-osc}, an xApp development framework~\cite{ricapp-osc}, and an E2 termination module~\cite{e2sim}. In parallel, open-source \gls{5g} \ran implementations such as \gls{oai}~\cite{oai} and srsRAN~\cite{srsran2023oran} have included native E2 terminations into their base station code bases. Together, these initiatives have fostered a novel ecosystem for experimenting with data-driven and \gls{ai}-enabled Open RAN solutions~\cite{bonati2023openran,johnson2021nexran,upadhyaya2023open}. 

Despite this progress, significant challenges remain. Developing xApps that interoperate seamlessly across heterogeneous components from different projects or vendors remains challenging, due to differences in software stacks and the complexity of low-level \gls{e2sm} procedures. As a result, many existing works restrict their validation to simulated environments or to a single \gls{ran} software stack, often considering only a limited set of \gls{ran} parameters. These drawbacks are further amplified when designing \gls{ai}-driven xApps, where developers must simultaneously focus on the control logic, handle \gls{e2sm} encoding and decoding, ensure portability across diverse \gls{ran} implementations, and address \gls{ai}-specific challenges such as data collection, model generalization, and validation. As a result, creating robust, interoperable \gls{ai}-driven xApps remains a non-trivial task, hindering the practical adoption of intelligence-driven closed-loop control in real-world deployments.

\textbf{Contribution.} In this paper, we propose \xdev, an xApp framework that enables intelligent closed-loop control of the \gls{ran} by streamlining xApp development. \xdev provides developers with a comprehensive \gls{sdk} offering high-level \glspl{api} that abstract the complexity of various \gls{e2sm} protocols. The framework supports seamless interaction between xApps, the \nearrt \gls{ric}, and the \gls{ran}-side E2 termination across different \gls{ran} software stacks. Thus, it allows the same xApp logic to operate concurrently across heterogeneous \gls{ran} implementations, enabling interoperability and consistent behavior even when different E2 nodes are connected to the same \nearrt \gls{ric}.

This work extends our previous monitoring-focused framework~\cite{feraudo2024xDevsm}, transforming \xdev into a control-enabling platform that provides the essential building blocks for applications that learn from observed metrics, predict network dynamics, and issue near-real-time control actions to optimize \gls{ran} performance. Beyond the framework improvements, this paper provides practical examples of control-based xApps and validates \xdev in real-world testbed experiments. We demonstrate the capabilities of \xdev through three representative use cases across heterogeneous \gls{ran} software stacks: (i) a \gls{kpm}-based monitoring scenario in which the xApp subscribes to \gls{ran} measurement reports, decodes them, and stores and visualizes the metrics for \gls{ai} and analytics workflows; (ii) slice-level \gls{prb} allocation using the \gls{rc} service model, applied to multiple \glspl{ue} and slices to enforce resource quotas; and (iii) a mobility- and bandwidth-aware closed-loop scenario, where \gls{kpm} telemetry guides \gls{rc}-based control actions for dynamic \gls{prb} allocation and handovers, demonstrating the seamless integration of monitoring and control in practical xApps.

The main contributions of this paper are summarized as follows:
\begin{itemize}
    \item \gls{ai}-Ready Control Framework: We design \xdev to support closed-loop control, enabling intelligent RAN optimization via programmable xApps. The framework provides standardized data pipelines (\gls{kpm}) and control interfaces (\gls{rc}) that naturally support reinforcement learning formulations, establishing a foundation for future AI-driven optimization.
    \item Practical control-based examples: We provide representative open-source xApps that demonstrate monitoring, resource allocation, and closed-loop control across multiple open-source \gls{ran} implementations, serving as concrete templates for developers building both conventional and \gls{ai}-driven solutions.
    \item Real-world validation: We experimentally validate \xdev across two real-world deployments, including a general-purpose multi-\gls{ran} setup and a large-scale testbed with distributed units and heterogeneous radio hardware, demonstrating its effectiveness in practical network conditions and its readiness for \gls{ai}/\gls{ml} integration.
\end{itemize}

The remainder of the paper is organized as follows. Section~\ref{sec:backintro} provides background on near-real-time \gls{ric} closed-loop control in O-RAN. Section~\ref{sec:related-work} discusses related work and the state of the art. Section~\ref{sec:xdevsm} presents the architecture of the enhanced \xdev framework. Section~\ref{sec:validation} presents the real-world validation of \xdev across multiple use cases. Finally, Section~\ref{sec:conclusions} concludes the paper.

\section{A Primer on Near-Real-Time Closed-Loop Control in O-RAN}
\label{sec:backintro}

\begin{figure}[h]
    \centering
    \includegraphics[width=\linewidth]{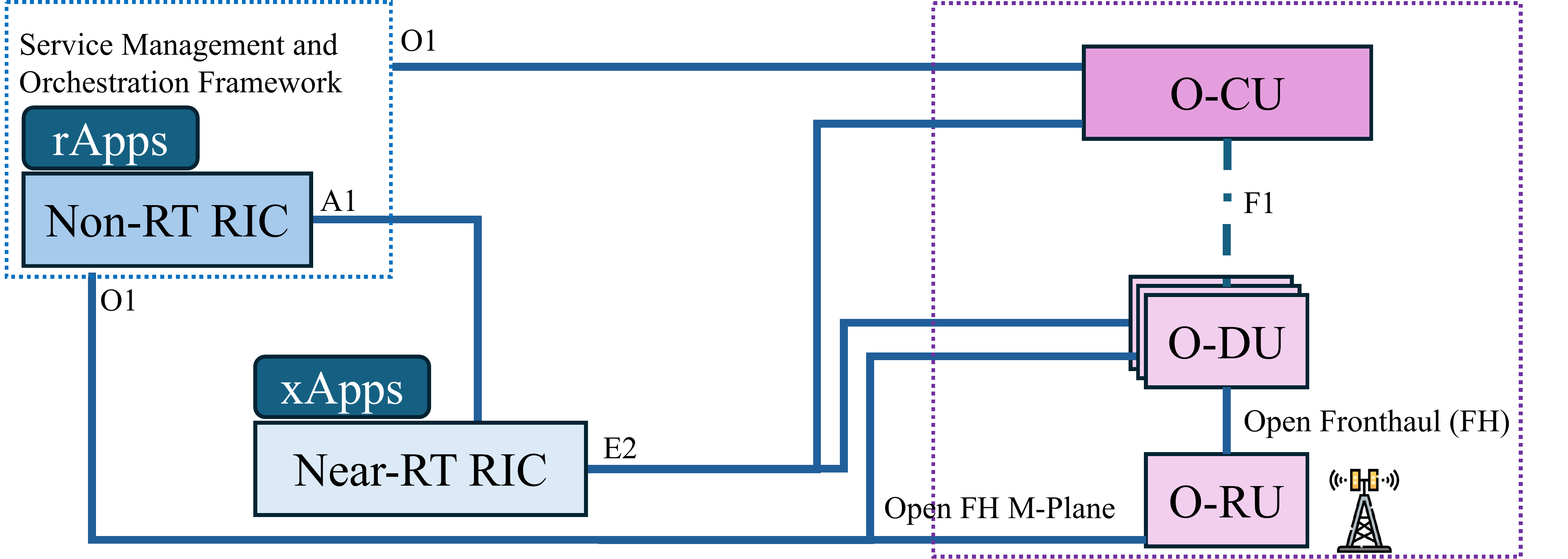}
    \caption{Overview of the O-RAN architecture and interfaces}
    \label{fig:oran}
\end{figure}

Figure~\ref{fig:oran} provides an overview of the O-RAN components and interfaces. In this paper, we focus specifically on the \nearrt closed-loop control of the \gls{ran}, emphasizing the characteristics of the \nearrt \gls{ric} and the E2 interface.
The \nearrt \gls{ric} enables control loops on a timescale between $10$\:ms and $1$\:s, through applications that support custom logic called xApps. To support these applications~\cite{oran-wg3-nrrt-ric, oran-wg3-e2-gap}, the \nearrt \gls{ric} provides (i)~a database called \gls{sdl} for storing \gls{ran} information; (ii)~a messaging infrastructure for inter-component communication; (iii)~termination for open interfaces and \glspl{api}; and (iv)~conflict management tools to resolve contention between xApps managing the same \gls{ran} resources. Each \nearrt \gls{ric} may be associated with multiple \gls{ran} functions---also called E2 nodes---via the E2 interface. This enables control over \gls{rrm} and other \gls{ran} functionalities.
%

The O-RAN ALLIANCE \gls{wg} 3 has defined the E2 interfaces through a series of documents and protocols~\cite{oran-wg3-e2-gap}. These documents specify the \gls{e2ap}, which defines the signaling and procedural protocols for coordination with E2 nodes~\cite{oran-wg3-e2-ap}, as well as the \gls{e2sm} formalizing \gls{ran} function definitions and services exposed by E2 nodes~\cite{oran-wg3-e2-sm}. Specifically, the \gls{e2ap} is used for general management operations, such as connection and disconnection of E2 nodes, and for passing application-specific messages between xApps and target \gls{ran} functions within an E2 node. The \gls{e2sm}, instead, describes the payload of \gls{e2ap} messages for different \glspl{sm}. A \gls{sm} expresses the semantics of interactions between xApps and E2 nodes. Each service model is described in a document that extends the general specifications of the \gls{e2sm} with parameters specific to the use cases of interest (e.g., metrics and the type of control supported).
%
When first establishing a connection, an E2 node shares its capabilities and supported \gls{ran} functions with the \nearrt \gls{ric}. These \gls{ran} functions are, then, advertised via the corresponding \gls{e2sm}. While the O-RAN ALLIANCE provides several documents for different \glspl{sm}~\cite{oran-wg3-e2-sm-kpmv3,oran-wg3-e2-sm-rc, oran-wg3-e2-sm-ccc, oran-wg3-e2-sm-llc}, the remainder of this section focuses on those supported by the \xdev framework, specifically, the \gls{kpm}~\cite{oran-wg3-e2-sm-kpmv3} and \gls{rc}~\cite{oran-wg3-e2-sm-rc} service models.

\subsection{E2 Service Model General Aspects}
\label{sec:genericsm}

\begin{figure}[h]
    \centering
    \includegraphics[width=\linewidth]{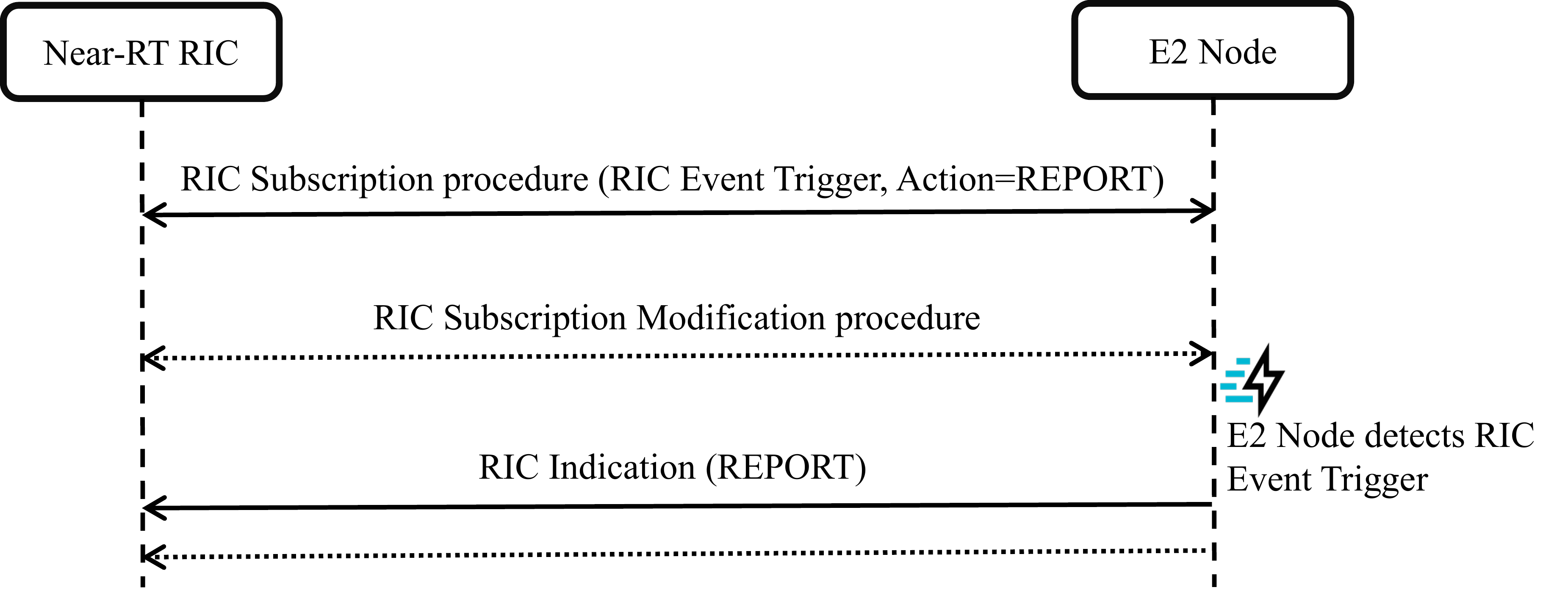}
    \caption{REPORT Subscription procedure}
    \label{fig:report}
\end{figure}

As defined by the O-RAN ALLIANCE \gls{wg}3~\cite{oran-wg3-e2-sm,oran-wg3-e2-gap}, each \gls{ran} function exposes a set of services over the E2 interface, which may include one or more of the following: REPORT, INSERT, CONTROL, POLICY, QUERY, and/or ASSISTANCE. These services are accessed via standardized \gls{e2ap} procedures, such as \gls{ric} Subscription, \gls{ric} Indication, and \gls{ric} Control.

The REPORT service is requested by the xApp running on the \nearrt \gls{ric} via a \gls{ric} Subscription procedure (see Fig.~\ref{fig:report}). During the subscription, the \nearrt \gls{ric} specifies the trigger and periodicity with which the E2 node should send reports. The INSERT service is similar to REPORT, but the E2 node does not continue the associated procedure; instead, it waits for a response from the \nearrt \gls{ric} or for a timer to expire. For example, a \gls{ric} Subscription may be defined with a trigger associated with an \gls{rrm} procedure, which is suspended when the insert message is sent.

\begin{figure}[h]
    \centering
    \includegraphics[width=\linewidth]{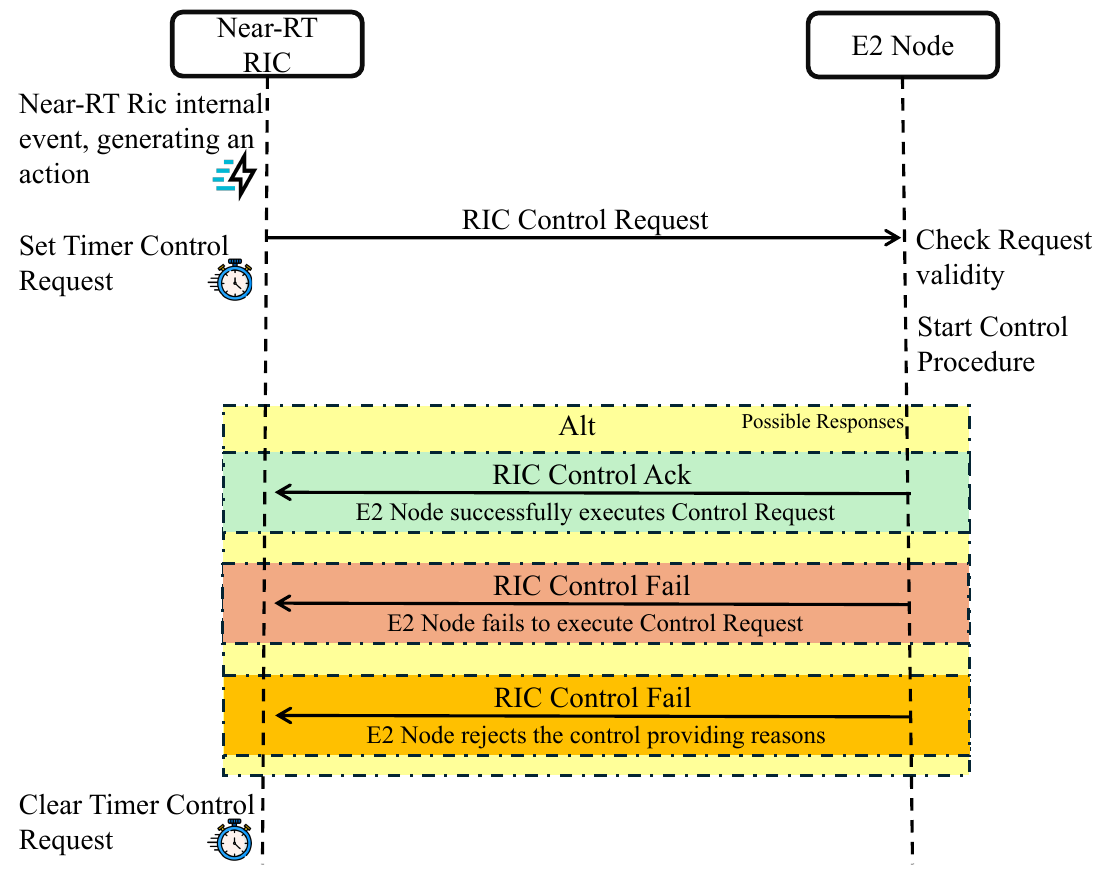}
    \caption{Control Request Procedure}
    \label{fig:control}
\end{figure}

The CONTROL service request (Fig.~\ref{fig:control}) is issued by an xApp on the \nearrt \gls{ric} to initiate or restart a control procedure associated with a specific control action defined in the request. The POLICY service instructs the E2 node to execute specific policies whenever the conditions described in the \nearrt \gls{ric}'s subscription request are met. The QUERY service allows the xApps to send requests to retrieve \gls{ran}-related and/or \gls{ue}-related information from the E2 node. The ASSISTANCE service, provided by the \nearrt \gls{ric}, allows E2 nodes to leverage control and support functionalities exposed by the \nearrt \gls{ric}. For instance, an E2 Node may send an ASSISTANCE REQUEST specifying the required service and any optional updates; the \nearrt \gls{ric} then responds with either the requested assistance or a failure cause. If updates are requested, the \gls{ric} may subsequently send ASSISTANCE INDICATION messages until the E2 Node explicitly halts the service.

To ensure consistency and interoperability, each \gls{e2sm} specifies a recommended method for encoding the information elements associated with a specific \gls{ran} function. 

\subsection{Key Performance Measurement Service Model}
The \gls{kpm} \gls{e2sm}~\cite{oran-wg3-e2-sm-kpmv3} enables an E2 node to periodically expose and transmit a set of measurements to the \nearrt \gls{ric}. As such, this \gls{sm} supports only the REPORT \gls{ric} service. For this service, five styles are defined, including \textit{E2 Node Measurement}, \textit{E2 Node Measurement for a single \gls{ue}}, \textit{Condition-based, \gls{ue}-level E2 Node Measurement}, \textit{Common Condition-based, \gls{ue}-level Measurement}, and \textit{E2 Node Measurements for multiple \glspl{ue}}. 

The procedure for initiating REPORT data streaming is illustrated in Fig.~\ref{fig:report}. An xApp running on the \nearrt \gls{ric} requests the metrics exposed by an E2 node and selects those relevant for its execution. Next, it initiates a \gls{ric} subscription procedure, specifying the conditions under which the E2 node sends measurements. These conditions include the \textit{RAN function ID} (set to 2 for KPM), the \textit{Event Trigger}, and the \textit{Action Definitions} (i.e., the specific measurements selected by the xApp~\cite{5gmeasurements}). Upon a successful subscription, the E2 node transmits the requested metrics to the xApp whenever the specified conditions are met. The metrics sent depend on those supported by the E2 node, whereas the triggering conditions are determined by the selected REPORT service style specified during subscription.


\subsection{RAN Control (RC) Service Model}
The \gls{rc} \gls{e2sm}~\cite{oran-wg3-e2-sm-rc} enables closed-loop control of E2 nodes. This service model is essential for radio resource management xApps. Unlike \gls{kpm}, the \gls{rc} \gls{sm} supports all the \gls{ric} services introduced in the previous section. However, in this paper, we focus exclusively on the CONTROL \gls{ric} service as it is currently the only one supported by the framework for this \gls{e2sm}.

To realize CONTROL operations, the \gls{rc} service model defines a set of control capabilities, referred to as CONTROL Service Styles. Among these, \xdev currently supports three styles: \gls{rbc}, \gls{rrac}, and \gls{cmmc}.

The \gls{rbc} style allows an xApp to modify radio bearers at the E2 node or for a specific \gls{ue}, such as reconfiguring the \gls{qos} profile of a \gls{drb}. The \gls{rrac} enables modification of resource allocation at various levels: E2 node, cell, slice, or \gls{ue}. For example, it includes a control action called \textit{Slice-level \gls{prb} quota}, which allows the xApp to influence the \gls{rrm} policy regarding \gls{prb} allocation per slice. The \gls{cmmc} style facilitates the initiation of handover procedures for connected \glspl{ue}, including actions such as \textit{handover control}, which triggers a handover between two cells for a given \gls{ue}.

In \gls{rc}-based scenarios, an xApp requires context and state information from the E2 node to apply control actions effectively. Therefore, such xApps often leverage multiple \glspl{sm}, e.g., using \gls{kpm} to gather measurement data and \gls{rc} to perform control, or rely on multiple \gls{ric} service procedures with RC as the underlying \gls{sm} (e.g., REPORT/INSERT in combination with CONTROL).

As shown in Fig.~\ref{fig:control}, once the xApp obtains the necessary information, it issues a \gls{ric} Control Request message to the E2 node. This message includes details of the control action, the procedure identification, and an optional acknowledgment request. 
If the control request is successfully executed, the E2 node responds with an acknowledgment (if requested). Otherwise, it sends a RIC Control Failure message indicating the reason for the failure.

\begin{table}[t]
\centering
\caption{Support of O-RAN E2 Service Models in Open-Source RAN Implementations. For \gls{rc} Control services, Action IDs are shown in parentheses; ``P" denotes partial support}
\label{tab:e2sm_support}
\renewcommand{\arraystretch}{1.2}
\resizebox{\columnwidth}{!}{%
\begin{tabular}{l|c|ccc|cc}
\hline
\textbf{RAN Stack} 
& \textbf{KPM} 
& \multicolumn{3}{c|}{\textbf{RC Control Service}} 
& \multicolumn{2}{c}{\textbf{RC Report Service}} \\
\cline{3-7}
& 
& \textbf{RBC} 
& \textbf{RRAC} 
& \textbf{CMMC} 
& \textbf{UE Info} 
& \textbf{Msg Copy} \\
\hline
\gls{oai}-2025 
& \checkmark 
& \checkmark\,(ID~1)P 
& -- 
& -- 
& \checkmark 
& \checkmark \\
\gls{oai} custom~\cite{cheng2024oranslice}
& \checkmark 
& \checkmark\,(ID~1)P 
& \checkmark\,(ID~6)
& \checkmark\,(ID~3)
& \checkmark 
& -- \\
srsRAN-2025 
& \checkmark 
& -- 
& \checkmark\,(ID~6) 
& \checkmark\,(ID~3) 
& -- 
& -- \\
\hline
\end{tabular}%
}
\end{table}

\section{Related Work}
\label{sec:related-work}

The growing adoption of open and programmable \glspl{ran} has driven extensive research into frameworks, tools, and methodologies that facilitate intelligent and interoperable control~\cite{polese2022understanding,giordani2020toward,chowdhury20206g}. Among the most comprehensive contributions to xApp development, the work in~\cite{santos2025managing} offers a detailed guide to deploying, developing, and managing xApps within \gls{osc}-based environments. Likewise, the study in~\cite{polese2022understanding} provides an in-depth analysis of O-RAN, covering its architectural principles, specifications, security aspects, and the \gls{ai}-driven closed-loop control capabilities it enables. 

Building on these foundations, a growing body of work has turned toward the design and validation of closed-loop optimization frameworks powered by \gls{ai} within the O-RAN ecosystem. For instance, Kouchaki \emph{et al.}~\cite{kouchaki2022actor} demonstrated how to develop and deploy a \gls{rl}-based xApp using the \gls{osc} framework, while Tsampazi \emph{et al.}~\cite{tsampazipandora} compared multiple deep \gls{rl} approaches for optimizing different \gls{ran} parameters, also leveraging the \nearrt \gls{ric}. Similarly, Qazzaz \emph{et al.}~\cite{qazzaz2025} implemented a machine-learning-based xApp for dynamic \gls{prb} allocation in a simulated environment.

Furthermore, recent works have demonstrated the practical applicability of \gls{ai}-driven algorithms in O-RAN deployments~\cite{polese2022colo}. At the \nonrt \gls{ric} layer, Staffolani \emph{et al.}~\cite{staffolani2024} introduced a proactive resource orchestrator based on \gls{rl} to learn traffic demand patterns and dynamically allocate resources. At the \nearrt \gls{ric}, the authors in~\cite{dzaferagic2024ml} validated an \gls{ml}-based handover prediction xApp using \gls{lstm}, while Vicario \emph{et al.}~\cite{vicario2025handover} proposed customized handover control mechanisms for intelligent mobility management. Other studies have addressed service-aware resource management, including baseband allocation and \gls{vnf} activation through network slicing~\cite{motalleb2023resource}.

To support these research directions, the O-RAN \gls{wg} 2 has released a specification~\cite{oran-wg2-ai-ml-workflow} that defines \gls{ai}/\gls{ml} lifecycle management procedures and criteria for selecting appropriate hosts for training and inference. Aligned with this effort, the \gls{osc} has begun developing an \gls{ai}/\gls{ml} framework comprising multiple interfaces and open-source modules to support the specification~\cite{aiframework-osc}.

Complementary to this, several works have proposed testbeds that enable the experimentation and validation of \gls{ai}-enabled \gls{ran} control. For example, Parada \emph{et al.}~\cite{PARADA2025101729} developed an open-source testbed for evaluating \gls{ai} techniques in O-RAN-compliant networks. In contrast, Upadhyaya \emph{et al.}~\cite{upadhyaya2023open} designed a software framework and testbed architecture that integrates \gls{ai} workloads within \nearrt \glspl{ric}. At the \gls{smo} layer, Polese \emph{et al.}~\cite{polese2025beyond} proposed an \gls{ai} orchestrator to manage resource allocation and introduced the concept of the \gls{ai} site for enabling real-time \gls{ai} execution within O-\gls{ran} environments. At the \gls{ran} layer, Gonzalez \emph{et al.}~\cite{gonzalez2024xstart} presented a framework for xApp development and validation within simulated ns-O-RAN environments~\cite{lacava2023ns}, whereas the authors in~\cite{flexapp} proposed a middleware solution enabling xApp portability across different \gls{ric} implementations using real implementations.

From a practical perspective, most real-world experimentation with O-RAN–compliant systems relies on a limited set of open-source \gls{ran} software stacks that currently support the E2 interface. In particular, \gls{oai}~\cite{oai}, and srsRAN~\cite{srsran} have emerged as the primary open-source reference implementations providing E2 connectivity with the \nearrt \gls{ric}. However, these platforms differ considerably in architecture, supported service models, and control capabilities, which complicates the development of portable xApps.

Table~\ref{tab:e2sm_support} provides an overview of the service models supported by the examined open-source RAN implementations, including a customized version of \gls{oai}. For the \gls{rc} Control and \gls{rc} Report services, the table specifies the individual control actions corresponding to each service style, with their Action ID indicated in parentheses. The symbol “P” denotes partial support, meaning that while the platform recognizes the service or action, it may not be fully enforced or operational in practice. The custom \gls{oai} version referenced in the table builds upon~\cite{cheng2024oranslice} and has been further extended with O-RAN-compliant \gls{rc} functionalities. This comparison emphasizes why many existing solutions either focus on a single \gls{ran} stack or rely on simulation-based setups, leaving cross-stack interoperability largely unaddressed.

Hence, despite these advancements, existing frameworks typically target specific use cases or rely on simulated environments, limiting their applicability and reproducibility. Moreover, most solutions do not fully leverage parameters defined by O-RAN-compliant service models for algorithm training, nor do they abstract the underlying \gls{ran} software complexity~\cite{srsran2023oran}, which hinders real-world integration. As a result, deploying portable, \gls{ai}-ready xApps that operate consistently across different \gls{ran} stacks remains a significant challenge.

\begin{figure}
    \centering
    \includegraphics[width=.6\linewidth]{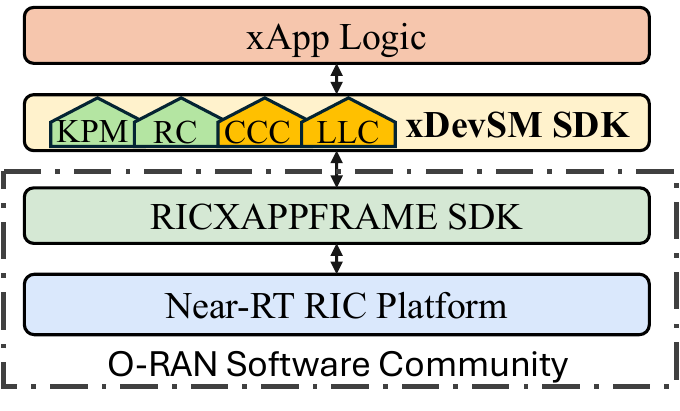}
    \caption{Overview of xApp stack using the xDevSM framework}
    \label{fig:xappsmframe}
\end{figure}

Among the existing efforts, the \gls{ai}/\gls{ml} framework proposed by the \gls{osc}~\cite{aiframework-osc} represents a valuable step toward integrating intelligent algorithms across the entire O-RAN architecture, spanning both non-\gls{rt} and near-\gls{rt} \gls{ric} layers. However, unlike the framework proposed in this work, the \gls{osc} framework does not explicitly target the development of xApps or the simplification of their interaction with \gls{e2sm}-based interfaces. Instead, it focuses on the broader \gls{ai} lifecycle covering training, inference, and deployment within the O-RAN stack. As such, the \gls{osc} \gls{ai}/\gls{ml} framework and \xdev are complementary: while the former enables end-to-end \gls{ai} integration, \xdev provides the tools and abstractions necessary to design, implement, and validate xApps efficiently on real \gls{ran} deployments.

It is also worth noting that an earlier version of \xdev was presented as a preliminary step toward simplifying xApp development~\cite{feraudo2024xDevsm}. While that initial effort demonstrated the feasibility of abstracting low-level O-RAN interfaces, its scope was limited by an exclusive reliance on the \gls{kpm} service model. As a result, xApps were confined to observability functions and could not directly influence \gls{ran} behavior. Moreover, the original design did not support the concurrent use of multiple \glspl{sm} within a single xApp. This capability is essential for \gls{ai}-driven closed-loop control, where monitoring and actuation must be tightly integrated. Extending the framework to support the \gls{rc} service model, therefore, required a substantial architectural redesign. This redesign included generalizing internal abstractions, extending data pipelines, and introducing flexible mechanisms for composing multiple service models within a single xApp. These changes enable the development of portable closed-loop xApps and allow their experimental validation across multiple open-source \gls{ran} implementations.

\begin{figure}
    \centering
    \includegraphics[width=1\linewidth]{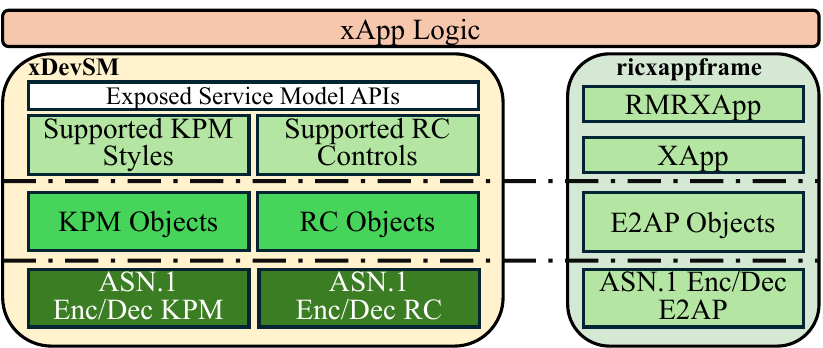}
    \caption{xDevSM modules overview}
    \label{fig:xdevsmmodules}
\end{figure}

\section{xDevSM: A Closed-Loop Control Framework for AI-Driven xApps}
\label{sec:xdevsm}

This section presents \xdev, a modular and developer-friendly framework for building O-RAN–compliant xApps for the \gls{osc} \nearrt \gls{ric}. The framework abstracts low-level \gls{e2sm} handling, allowing developers to focus on xApp logic through extensible, high-level \glspl{api}.\footnote{\url{https://github.com/wineslab/xDevSM/tree/dev}}As illustrated in Fig.~\ref{fig:xappsmframe}, \xdev is built on top of the Python-based xApp framework provided by \gls{osc}~\cite{oranscframepy,oscxappframe}, namely \textit{ricxappframe}, and exposes service-model–specific \glspl{api} to xApp developers.

The framework currently supports closed-loop control through two \glspl{e2sm}, namely \gls{kpm} and \gls{rc} (highlighted in green in the figure). Specifically, \xdev allows multiple service models to be combined within a single xApp, enabling the joint use of measurement, configuration, and control interfaces. As a result, the framework extends beyond monitoring-only functionality and provides a full-fledged platform for issuing near–real-time control actions based on observed network metrics. By supporting multi–service-model xApps, \xdev establishes a foundation for \gls{ai}-driven control applications, enabling learning-based algorithms to autonomously optimize and manage \gls{ran} behavior.

In addition to these capabilities, \xdev integrates a complete \gls{ci}/\gls{cd} workflow for automated deployment and validation of xApps across multiple \gls{ran} stacks. This pipeline streamlines the building, deployment, and testing of \glspl{sm} and xApps across heterogeneous environments, ensuring consistency, reducing integration time, and providing comprehensive logs of both data-plane performance and control-plane interactions.


\subsection{Generic xApp Model}
As illustrated in Fig.~\ref{fig:xdevsmmodules}, \xdev exposes service-model–specific \glspl{api} that are dynamically associated with an xApp based on the \gls{sm} it employs. When developing an xApp with \xdev, the developer first instantiates a generic xApp, which serves as the common foundation and is subsequently extended with the required \gls{sm} functionalities according to the target use case.
This generic abstraction is provided by the \texttt{xDevSMRMRXapp} class, which extends the \texttt{RMRXapp} class from the \gls{osc} \textit{ricxappframe}. The \texttt{xDevSMRMRXapp} class encapsulates all initialization procedures required to deploy the xApp as a microservice within the \gls{osc} \nearrt \gls{ric} cluster. In addition, it offers a set of core \glspl{api} that expose xApp metadata and E2 node–related information, independent of any specific service model.

Once an instance of this generic xApp is created, developers can incrementally enrich it by attaching one or more service-model modules, thereby enabling monitoring, control, or combined closed-loop functionalities within the same xApp logic.

\subsection{Monitoring}
The framework provides a dedicated class for the \gls{kpm} service model, \texttt{XappKpmFrame}, which exposes high-level \glspl{api} for interacting with E2 nodes. 
These include retrieving node-specific information, such as the list of supported \gls{kpm}-based \gls{ran} functions, and managing subscription procedures. In particular, it allows developers to access decoded \gls{ran} function definitions~\cite{oran-wg3-e2-sm} related to the \gls{kpm} service model, enabling them to select the desired functions for telemetry collection. An example of these \gls{ran} function definitions, as supported by a monolithic \gls{gnb} and exposed via \xdev, is shown in Listing~\ref{lst:kpmranfunc}. 

\begin{lstlisting}[float=t,language=logs,style=mystile-logs,
caption={KPM RAN Function Definitions extracted by xDevSM},
label={lst:kpmranfunc}]
{..."msg": "Available functions: {0: [], 1: [], 2: [], 3: ['DRB.PdcpSduVolumeDL', 
 'DRB.PdcpSduVolumeUL', 'DRB.RlcSduDelayDl', 'DRB.UEThpDl', 'DRB.UEThpUl', 
 'RRU.PrbTotDl', 'RRU.PrbTotUl'], 4: []}"}
\end{lstlisting}

Based on this information, developers can initiate a subscription using the \texttt{subscribe()} method provided by the framework, thereby enabling the reception of \gls{kpm}-related event reports. This method supports a range of configuration parameters, including the reporting interval, \gls{snssai} (SST and SD), action types (see Section~\ref{sec:backintro}), and more. At the time of writing, \xdev supports the \textit{Common Condition-based, UE-level Measurement} \gls{kpm} style. Consequently, only actions and parameters associated with this specific \gls{kpm} style can be encoded during the subscription procedure.

The encoding and decoding of \gls{kpm} messages are handled by intermediate modules, illustrated in Fig.~\ref{fig:xdevsmmodules} (\gls{kpm} Objects). These modules consist of \gls{kpm}-specific Python objects that wrap the C data structures provided by shared libraries generated using the FlexRIC toolchain~\cite{flexric}. As shown by the dark green blocks in the figure, these libraries encapsulate low-level ASN.1 processing and message serialization, thereby abstracting the complexity of the underlying E2 protocol stack from xApp developers.

To build a \gls{kpm}-based xApp, developers first create an instance of the generic \texttt{xDevSMRMRXapp} class, described in the previous subsection. To enrich this base xApp with \gls{kpm}-specific capabilities, they then instantiate the \texttt{XappKpmFrame} class, passing the \texttt{xDevSMRMRXapp} instance as an argument. This modular design allows domain-specific functionality to be attached to a generic xApp by composing it with the desired service model. An example of an xApp extended with \gls{kpm} capabilities is provided in Listing~\ref{lst:kpmex}.

\begin{lstlisting}[float=t, style=mypython, caption={Creating a KPM xDevSM xApp},
label={lst:kpmex}]
# Creating a generic xDevSM RMR xApp
xapp_generic = xDevSMRMRXapp(...)

# Adding kpm functionalities to the xapp
kpm_func = XappKpmFrame(xapp_generic, other_params)

# getting ran function description
kpm_func.get_ran_function_description(...)
\end{lstlisting}
\begin{lstlisting}[float=t, style=mypython, caption={Creating a RC xDevSM xApp},
label={lst:rcex}]
# Creating a generic xDevSM RMR xApp
xapp_generic = xDevSMRMRXapp(...)

# Adding rc functionalities to the xapp
rc_func = RadioResourceAllocationControl(xapp_gen, 
control_based_params)
\end{lstlisting}

\subsection{Control}

As outlined in Section~\ref{sec:backintro}, \xdev supports three control service styles of the \gls{rc} service model, each associated with one control action. Specifically, it supports: (i) \textit{QoS flow mapping configuration} as part of the \gls{rbc}, (ii) the \textit{slice-level \gls{prb} quota} as a \gls{rrac} action, and (iii) \textit{handover control} as a \gls{cmmc} action. The details of these control actions are provided in Section~\ref{sec:backintro}.

The framework provides a base class, \texttt{RCControlBase}, which defines the generic behavior of an \gls{rc}-based xApp. This provides \glspl{api} to get \gls{ran} function definition related to the \gls{rc} \gls{sm} and build \gls{ue} related information based on the E2 node deployment.  A dedicated class extends this for each service style: \texttt{RadioBearerControl}, \texttt{RadioResourceAllocationControl}, and \texttt{ConnectedModeMobilityControl}. Each of these classes encapsulates the parameters and procedures mandated by the O-\gls{ran} specification, while allowing developers to issue control requests through a unified \gls{api}.

As for the \gls{kpm} case, the framework provides \gls{rc}-related Python objects that wrap the C data structures provided by the shared libraries (dark green blocks in Fig.~\ref{fig:xdevsmmodules}). These contain methods that build the Control Request according to the service style selected for the specific use case. The developer can add new service styles by creating a new subclass of \texttt{RCControlBase} and adding the corresponding method in the \gls{rc} objects that builds the control request.

Similar to the \gls{kpm} case, developers who wish to build an \gls{rc}-based xApp must extend the base xApp with the required \gls{rc} functionalities. An example of such an instantiation is provided in Listing~\ref{lst:rcex}.

\subsection{Toward AI-Driven xApps: \xdev for Monitoring and Control Integration}
While \xdev allows developers to build xApps that rely on individual service models, its main strength lies in enabling the seamless integration of multiple models within a single application. In particular, combining the \gls{kpm} and \gls{rc} service models enables developers to implement closed-loop control applications, in which network measurements obtained via \gls{kpm} subscriptions directly inform control decisions executed via \gls{rc} actions.

As illustrated in Listing~\ref{lst:kpmrc}, an xApp can subscribe to \nearrt performance indicators, such as \gls{prb} utilization or QoS-related metrics, using the \texttt{XappKpmFrame}. Based on the observed telemetry, the same xApp can then dynamically adjust slice-level resource allocations through the \texttt{RadioResourceAllocationControl} class. Specifically, by incorporating both \gls{kpm} and \gls{rc} functionalities, the xApp first establishes subscription to the relevant \gls{kpm} metrics via the \texttt{subscribe} \gls{api}, and next issues CONTROL requests via the \gls{rc}-based \glspl{api}, according to the application logic.

\begin{lstlisting}[float=t, style=mypython, caption={Combining KPM monitoring and RC control in a single xApp},
label={lst:kpmrc}]
# Create a generic xDevSM RMR xApp
xapp = xDevSMRMRXapp(...)

# Add RC control capabilities
rc_func = RadioResourceAllocationControl(xapp, ...)

# Add KPM monitoring capabilities
kpm_func = XappKpmFrame(rc_func, ...)

# Subscribe to Metrics
kpm_func.subscribe(...)

def handle(...):
    # Define control behavior based on metrics
  
\end{lstlisting}

This tight integration of monitoring and control capabilities enables adaptive and intelligent orchestration of the \gls{ran}, aligning with the O-RAN vision of automated, data-driven network management and paving the way for \gls{ai}-driven closed-loop control applications.

\subsection{Automated Deployment and CI/CD Support}

\begin{figure}[t]
    \centering
    \includegraphics[width=0.9\columnwidth]{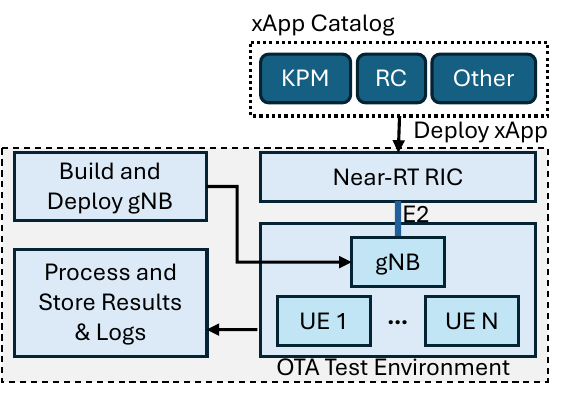}
    \caption{CI/CD pipeline}
    \label{fig:cicd-pipeline}
    \vspace{-.5cm}
\end{figure}

\begin{figure*}[t]
    \centering
    \begin{subfigure}[b]{0.48\linewidth}
        \centering
        \includegraphics[width=\linewidth]{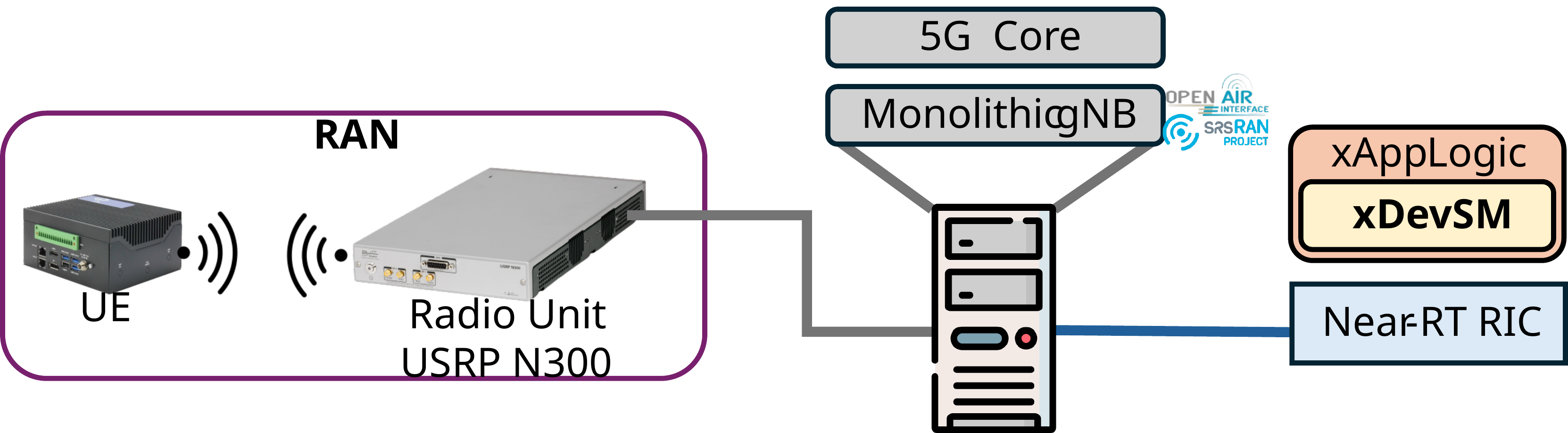}
        \caption{Single UE Deployment}
        \label{fig:singleUE}
    \end{subfigure}
    \hfill
    \begin{subfigure}[b]{0.48\linewidth}
        \centering
        \includegraphics[width=\linewidth]{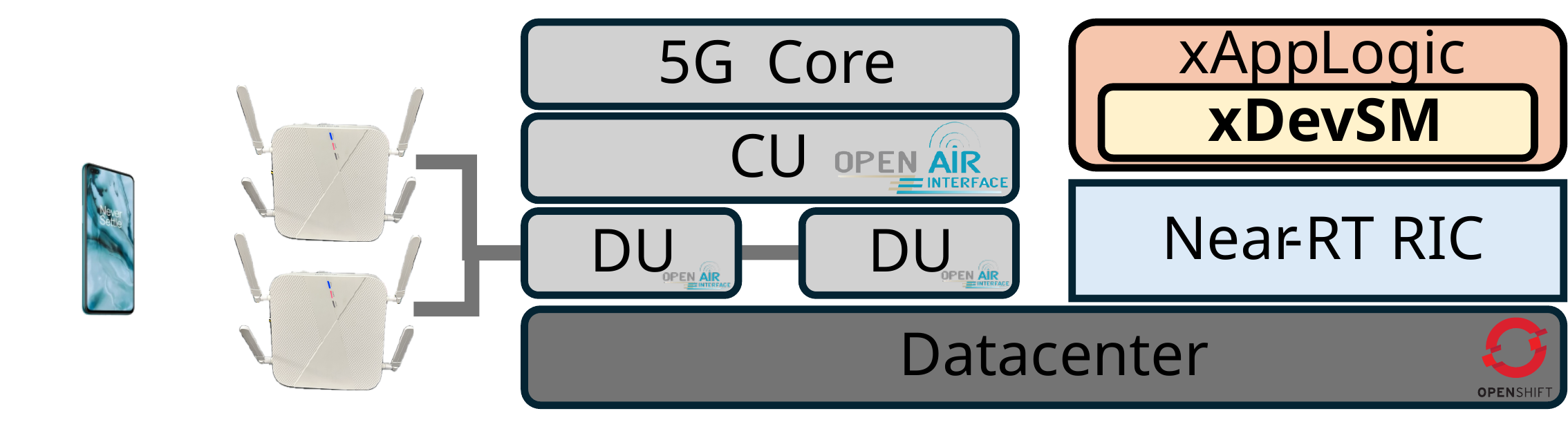}
        \caption{Two DUs Deployment}
        \label{fig:dus}
    \end{subfigure}
    \caption{Real-world Deployments}
    \label{fig:deployments}
\end{figure*}

Finally, we integrated a \gls{ci}/\gls{cd} pipeline into the \xdev framework, shown in Fig.~\ref{fig:cicd-pipeline}. The pipeline automates steps such as building \gls{gnb} container images using the \glspl{sm} provided by \xdev and deploying the resulting \gls{gnb} image as a software container on a Red Hat OpenShift compute cluster (see also Fig.~\ref{fig:dus}).
After the deployment step, the \gls{gnb} is automatically tested in an \gls{ota} environment, where it connects to a \nearrt \gls{ric}, and serves traffic to mobile \glspl{ue} (e.g., \gls{cots} \glspl{ue}, 5G modems, etc.).
During the automatic test, relevant \xdev xApps (e.g., \gls{kpm} xApp, \gls{rc} xApp) are deployed from an xApp catalog to the \nearrt \gls{ric} and interface with the \gls{gnb} under testing, for instance, requesting metrics available at the \gls{gnb} or sending control actions.
After the test completes, results and logs are processed and stored for troubleshooting and analysis operations performed by the developers. 
The pipeline generates three categories of logs: (i) Test summary Cell logs, which capture cell-level metrics and performance indicators; (ii) Test summary \gls{ue} logs, which record per-\gls{ue} connection statistics; and (iii) E2 Agent logs, which track the communication between the \gls{gnb}'s E2 termination point and the \nearrt \gls{ric}, including \gls{sm} subscription requests, indication messages, and control commands.
These logs provide comprehensive visibility into both data-plane performance and control-plane interactions, enabling developers to identify issues across different layers of the \gls{ran} stack.

%
The adoption of this \gls{ci}/\gls{cd} pipeline demonstrates that \xdev is not limited to one-off experiments but supports a complete DevOps workflow for \gls{ran} intelligence. This continuous process ensures that new \glspl{sm} and xApps, or their updates, can be automatically validated and deployed across heterogeneous environments, accelerating research and reducing integration time and complexity.

\begin{figure}
    \centering
    \includegraphics[width=\linewidth]{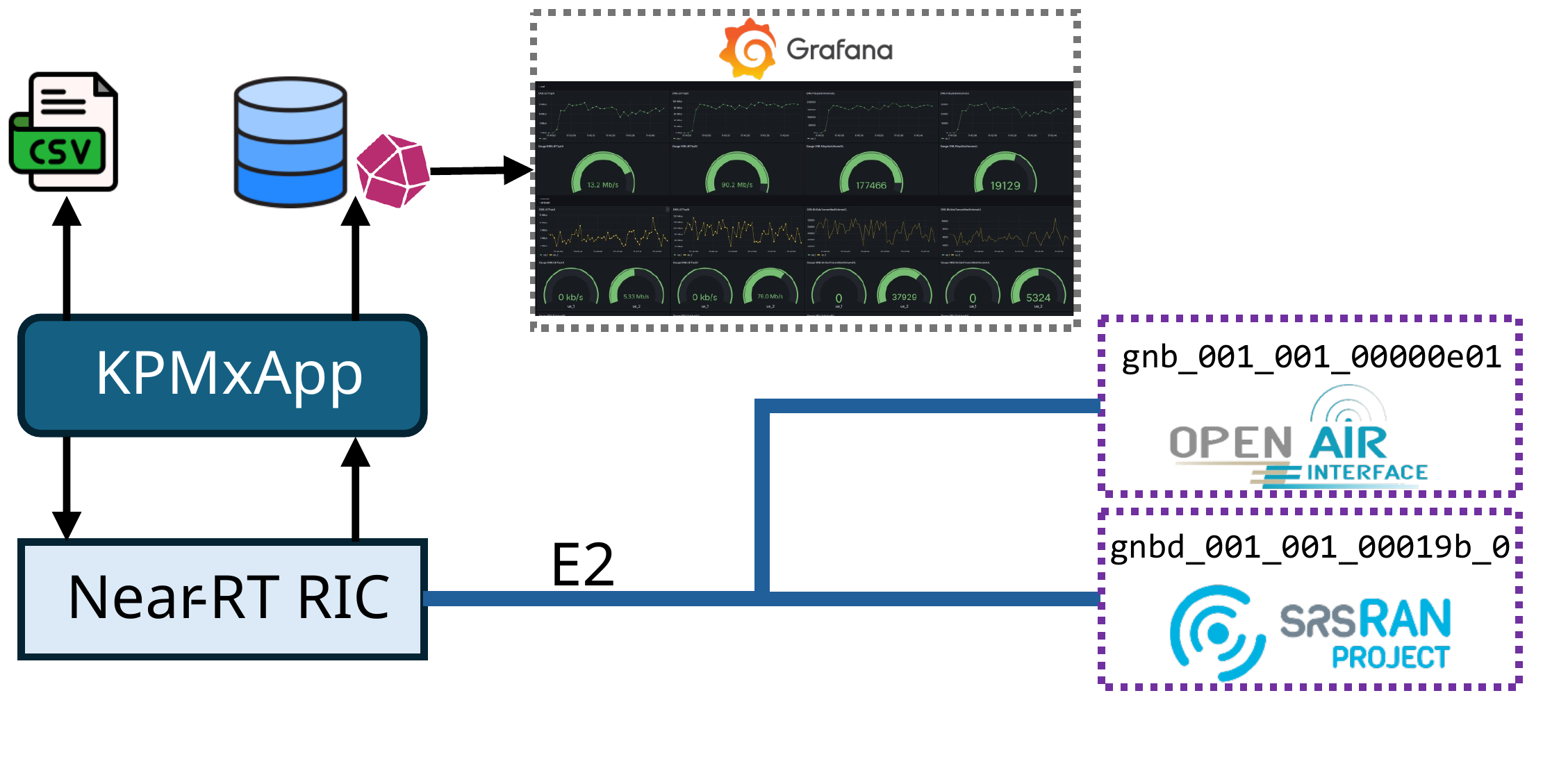}
    \caption{RAN Monitoring Use Case}
    \label{fig:usecase1}
\end{figure}
\begin{figure*}
    \centering
    \begin{subfigure}[b]{0.48\textwidth}
        \centering
        \resizebox{\linewidth}{!}{\input{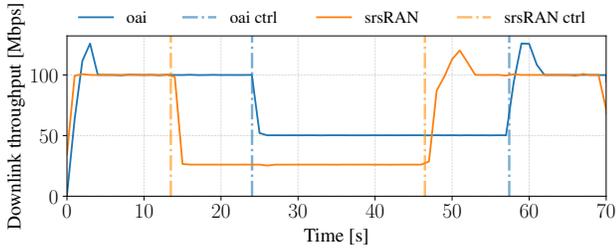}}
        \caption{Slice-Level PRB Control on srsRAN and OAI}
        \label{fig:prb1}
    \end{subfigure}
    \hfill
    \begin{subfigure}[b]{0.48\textwidth}
        \centering
        \resizebox{\linewidth}{!}{\input{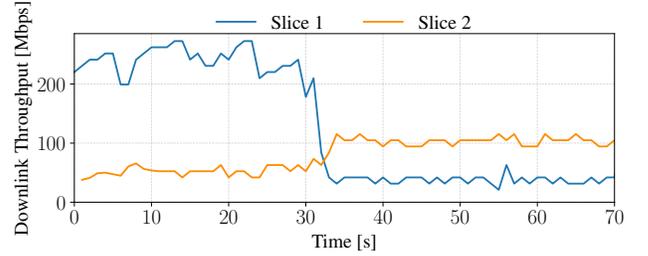}}
        \caption{Multi-slices PRB Control}
        \label{fig:prb2}
    \end{subfigure}
    \caption{Radio Resource Allocation Control}
    \label{fig:prbAll}
\end{figure*}

\section{Closed-Loop Control Use Cases and Validation Across Multiple RAN Implementations with xDevSM}
\label{sec:validation}
This section evaluates \xdev using a comprehensive set of experimental use cases on real-world testbeds that integrate heterogeneous open-source \gls{ran} implementations. Rather than focusing on isolated control primitives, we validate the framework by implementing complete monitoring, control, and closed-loop xApps that exercise different combinations of \gls{kpm}-based telemetry and \gls{rc}-based control actions. The experiments cover \gls{ran} monitoring, slice-aware radio resource allocation, adaptive closed-loop control, and mobility management, and are executed across multiple E2-enabled \gls{ran} stacks. Through these deployments, we demonstrate that \xdev enables portable and reusable xApp logic that operates across diverse E2 nodes, while abstracting differences in service-model support and \gls{ran}-specific control implementations.

\subsection{Experimental Setups}
The validation of \xdev was conducted on two complementary experimental deployments, designed to jointly assess monitoring, control, and closed-loop capabilities across heterogeneous \gls{ran} implementations. The first setup supports all slice-aware monitoring and control experiments across multiple \gls{ran} stacks, while the second is specifically tailored to validate mobility control in a multi-\gls{du} scenario.

The first deployment, illustrated in Fig.~\ref{fig:singleUE} consists of a single \gls{ue} connected to an E2-enabled \gls{gnb} running on a general-purpose PC. The \gls{gnb} interfaces with a \gls{usrp} N300 \gls{sdr} and operates in the $3.9$ GHz n77 band with a $40$ MHz bandwidth. The E2 node is connected to the \gls{osc} \nearrt \gls{ric}, which is deployed on a virtual machine running Kubernetes within a Proxmox cluster. This setup supports both \gls{oai} and srsRAN, enabling a direct comparison of monitoring and control behavior across different open-source \gls{ran} stacks under identical radio conditions.

The second deployment, shown in Fig.~\ref{fig:dus}, leverages the X5G~\cite{villa2025x5g} testbed and the AutoRAN framework~\cite{maxenti2025autoranautomatedzerotouchopen}, and is based on the NVIDIA ARC-OTA software stack. The \gls{ran} comprises two Foxconn RPQN \glspl{ru}, connected to two distributed units and a shared centralized unit, enabling F1-based inter-\gls{du} handovers. The system operates at a central frequency of $3.75$\,GHz with bandwidths of $40$\,MHz and $100$\,MHz. In this deployment, the \nearrt \gls{ric} is deployed as a set of OpenShift pods within a datacenter. Due to current software constraints, only \gls{oai} is used in this setup, as it is the only open-source \gls{ran} stack compatible with the NVIDIA ARC framework.

Both deployments use Open5GS as the core network. This dual-testbed approach allows us to validate \xdev across diverse operational conditions, ranging from single-cell slice control to multi-\gls{du} mobility management.

\subsection{Discussion on Limitations of Current RAN Stacks}

Before presenting the experimental validation, it is important to clarify the current limitations of open-source \gls{ran} implementations with respect to O-RAN control capabilities. As summarized in Table~\ref{tab:e2sm_support}, the \gls{rbc} service style is only partially supported by \gls{oai} at the time of writing. Specifically, the stack correctly parses and acknowledges \gls{rbc} CONTROL requests issued by the xApp, but the corresponding actions are not yet enforced within the \gls{ran} scheduler.

As a result, our evaluation for this service style is limited to validating the correctness of message encoding, transmission, and reception over the E2 interface, without assessing the impact of the control actions on radio behavior. Although this prevents a full end-to-end performance analysis, it still provides a useful validation of the \xdev abstractions and \glspl{api} for developing \gls{rbc}-based xApps. For completeness and reproducibility, we release the implementation of this xApp\footnote{https://github.com/wineslab/xDevSM-xapps-examples/tree/dev/radio\_bearer\_control\_xapp} as a reference example, illustrating how such controls can be realized using \xdev. We expect that a complete experimental validation will become possible as soon as full \gls{rbc} enforcement is integrated into open-source \gls{ran} stacks.

\subsection{RAN Monitoring and Telemetry Collection}
As a first validation step, we implement a monitoring xApp\footnote{\url{https://github.com/wineslab/xDevSM-xapps-examples/tree/dev/kpm_basic_xapp}} using the \texttt{XappKpmFrame} class provided by \xdev, which subscribes to \gls{kpm}-based measurements exposed by E2 nodes, specifically \gls{oai} and srsRAN \glspl{gnb}. As illustrated in Fig.~\ref{fig:usecase1}, the xApp decodes the received telemetry and stores it in InfluxDB, enabling real-time visualization through Grafana as well as offline access for further analysis. The collected data can also be exported in CSV format, facilitating integration with external processing pipelines or \gls{ai}-based workflows.
\begin{figure*}
    \centering
    \begin{subfigure}[b]{0.48\textwidth}
        \centering
        \resizebox{\linewidth}{!}{\input{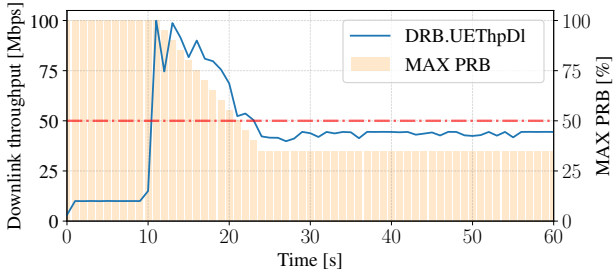}}
        \caption{KPM and PRB metrics in OAI}
        \label{fig:oaikpmprb}
    \end{subfigure}
    \hfill
    \begin{subfigure}[b]{0.48\textwidth}
        \centering
        \resizebox{\linewidth}{!}{\input{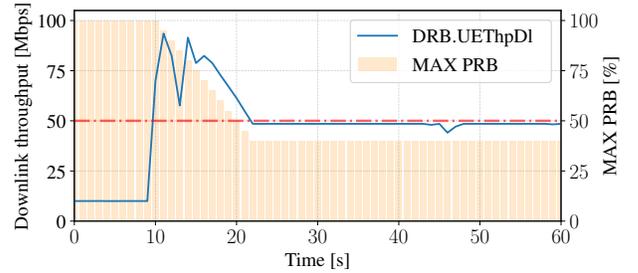}}
        \caption{KPM and PRB metrics in srsRAN}
        \label{fig:srskpmprb}
    \end{subfigure}
    \caption{Dynamic PRB Allocation Using KPM-RC xApp}
    \label{fig:usecase3}
\end{figure*}
The monitoring pipeline operates as follows: (i) the \nearrt \gls{ric} hosts the \gls{kpm} xApp, which manages subscription procedures and decodes incoming reports; (ii) the \glspl{gnb} (\gls{oai} and srsRAN) expose heterogeneous sets of performance metrics through the \gls{kpm} service model; and (iii) a data handling layer stores, visualizes, and optionally exports the collected measurements.

Unless explicitly stated otherwise, the performance indicators and time-series data shown in the following figures are gathered through this monitoring xApp. This use case highlights how \xdev abstracts the complexity of managing multiple \gls{kpm} subscriptions. It harmonizes heterogeneous \gls{gnb} outputs and interfaces with storage and visualization tools. As a result, it provides a reusable monitoring component that supports the subsequent closed-loop control experiments.

\subsection{Slice-Aware Radio Resource Allocation Control}
Building on the monitoring capabilities introduced above, we next validate \xdev in a control-oriented scenario by implementing a slice-aware Radio Resource Allocation Control xApp.\footnote{\url{https://github.com/wineslab/xDevSM-xapps-examples/tree/dev/prb_control_xapp}} This xApp leverages the \textit{slice-level \gls{prb} quota action} to dynamically regulate the percentage of physical resource blocks allocated to a given network slice.
As introduced in Section~\ref{sec:backintro}, this control is expressed through three \gls{rrm} parameters: \textit{dedicated}, \textit{minimum}, and \textit{maximum} ratios. In our experiments, the \textit{dedicated} and \textit{minimum} ratios are fixed to 5\% and 20\%, respectively, while the \textit{maximum} ratio is varied to assess its impact on \gls{ue} performance.

The experiment is conducted using the deployment in Fig.~\ref{fig:singleUE}, using srsRAN and the custom version of \gls{oai} (see Table~\ref{tab:e2sm_support}). In both cases, the controlled slice is identified by an \gls{snssai} with \gls{sst}~1 and \gls{sd}~1.

Figure~\ref{fig:prb1} reports the results collected via the \gls{kpm} xApp
during a downlink UDP transmission of 100~Mbps. At the instants labeled \texttt{srs ctrl} and \texttt{oai ctrl}, the xApp limits the maximum \gls{prb} allocation to 40\%, before restoring it to 100\%. In both \gls{ran} stacks, this results in a clear and immediate reduction in throughput, followed by a recovery, confirming correct enforcement of the control action. From these two control exchanges performed during the experiment, we also measure the reaction time of the respective \gls{ran} software stacks when processing control messages. The time between sending a control request and receiving the acknowledgement (see Fig.~\ref{fig:control}) is, on average, 4.5 ms for the \gls{oai} software stack and 7 ms for the srsRAN software stack.

To further validate slice-level control in a more realistic multi-slice scenario, we apply the same control logic to a deployment with two concurrent network slices hosted by a single \gls{gnb}. This experiment is conducted on the setup shown in Fig.~\ref{fig:dus}, but using a single \gls{du} operating at $40$ MHz and two \glspl{ue}: a COTS device (OnePlus AC2003) attached to Slice~1 and a software-based \gls{ue} attached to Slice~2. Both \glspl{ue} generate downlink iperf3 traffic, with the COTS device starting first.
As shown in Fig.~\ref{fig:prb2}, the scheduler initially allocates a larger share of \glspl{prb} to Slice~1. When the xApp enforces a maximum \gls{prb} cap of 30\% on Slice~1, its throughput is immediately reduced, allowing Slice~2 to acquire additional radio resources. The throughput of the software-based \gls{ue} increases accordingly, up to its hardware-imposed limit of approximately 100~Mbps.

Building on the previous experiments, we finally demonstrate a fully closed-loop xApp\footnote{\url{https://github.com/wineslab/xDevSM-xapps-examples/tree/dev/kpm_prb_xapp}} that jointly leverages \gls{kpm}-based monitoring and \gls{rc}-based control to enforce a throughput constraint on a network slice. The xApp continuously monitors slice-level throughput metrics via \gls{kpm}. When the measured throughput exceeds a predefined threshold, it issues a \textit{slice-level \gls{prb} quota action} that reduces the maximum \gls{prb} allocation by 5\% at each control step until the constraint is satisfied.

This experiment is conducted using the deployment in Fig.~\ref{fig:singleUE}, with a single \gls{ue} and E2 node, and is validated on both \gls{oai} and srsRAN. The \gls{ue} generates two sequential downlink iperf3 flows: a 10~Mbps flow for the first 10~s, followed by a 100~Mbps flow for the next 50~s. The throughput threshold enforced by the xApp is set between these two values.

Figure~\ref{fig:usecase3} shows that, once the offered traffic exceeds the threshold, the xApp reacts by progressively tightening the \gls{prb} allocation. With \gls{oai} (Fig.~\ref{fig:oaikpmprb}), the throughput stabilizes below the threshold when the maximum \gls{prb} allocation reaches 35\%, whereas with srsRAN (Fig.~\ref{fig:srskpmprb}) this occurs at 40\%. Although both stacks operate with the same number of available \glspl{prb}, these differences arise from stack-specific \gls{kpm} reporting granularity and scheduler behavior. The srsRAN stack exhibits smoother throughput adaptation, while \gls{oai} shows larger short-term fluctuations under tight \gls{prb} constraints.

\subsection{Mobility-Aware RAN Optimization and Control}
We next evaluate \xdev’s ability to manage mobility-related actions. In this scenario, we focus on the \textit{handover control} action of the \gls{cmmc} style, which orchestrates primary cell handovers to a target cell. As previously anticipated, the deployment used is shown in Fig.~\ref{fig:dus}.

\begin{figure}
    \centering
    \resizebox{\columnwidth}{!}{\input{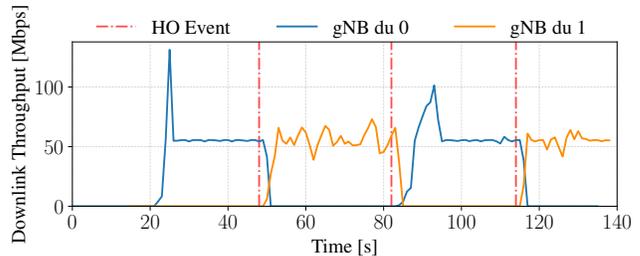}}
    \caption{F1 Handover two DUs one UE}
    \label{fig:hocontrol}
\end{figure}

In this case, the \xdev xApp\footnote{\url{https://github.com/wineslab/xDevSM-xapps-examples/tree/dev/ho_xapp}} collects \gls{ue}-related information through the \gls{kpm} service model, using it to determine which \gls{ue} should be handed over. Based on these measurements, the xApp issues a \textit{handover control} request to transfer the \gls{ue} from one \gls{du} to the other within the same \gls{gnb}. During the experiment, the \gls{ue} generates a 50~Mbps downlink TCP iperf3 flow, while the xApp sends the handover command to the \gls{cu}, which triggers the F1-based inter-\gls{du} handover.
We execute this control loop three times, indicated by the red vertical lines in Fig.~\ref{fig:hocontrol}. The results show corresponding shifts in measured traffic between the two \glspl{du}, demonstrating that the \gls{ran} correctly executes the handover procedure initiated by the xApp. This experiment complements the previous monitoring and slice-level control use cases, showing that \xdev can integrate telemetry from \gls{kpm} to drive both resource allocation and mobility management in heterogeneous \gls{ran} deployments.

\section{Conclusions and Future Work}
\label{sec:conclusions}
This work presents \xdev, an enhanced framework that enables \gls{ai}-\gls{ran} closed-loop control xApps Across Heterogeneous \gls{ran} software stacks. Through a modular architecture and unified \glspl{api}, \xdev allows developers to focus exclusively on xApp logic while the framework handles message encoding/decoding, subscription management, and service-model composition.

We validate \xdev on real-world testbeds using commercial \glspl{ue}, \gls{usrp}-based \glspl{sdr}, and Foxconn radio units, demonstrating seamless interoperability across both \gls{oai} and
srsRAN \glspl{gnb}. To showcase its capabilities, we implemented multiple \gls{rc} control loops, including \gls{rrac} and \gls{cmmc}, and ensured reliability through a full \gls{ci}/\gls{cd} pipeline. We also developed xApps that cover key O-RAN scenarios: one for network performance monitoring using \glspl{kpm}, one for slice-level \gls{prb} allocation across multiple \glspl{ue} and slices, and one for mobility-aware handover control. These experiments demonstrate that \xdev enables intelligent closed-loop applications combining monitoring and control across heterogeneous \gls{ran} deployments.

\subsection{Toward AI-Driven RAN Optimization}

While the experimental validation in this paper focuses on rule-based and threshold-based control policies, \xdev architecture is explicitly designed to support \gls{ai} and machine learning workflows. The framework provides the essential infrastructure that makes \gls{ai}-driven \gls{ran} optimization feasible in real-world deployments:
\begin{itemize}
    \item Standardized Data Pipelines: \xdev abstracts away the complexity of \gls{kpm} subscriptions and metric decoding, delivering structured measurements (throughput, \gls{prb} utilization, latency, etc.) that can be directly fed into \gls{ml} pipelines. The integration with InfluxDB and CSV export enables offline training, while real-time metric streaming supports online learning scenarios.
    \item Control Abstraction as Action Space: \xdev high-level control \glspl{api} abstract \gls{e2sm} encoding details, allowing \gls{ai} algorithms to issue control commands through simple function calls rather than managing low-level protocol messages.
    \item Safe Exploration and Deployment: The \gls{ci}/\gls{cd} pipeline supports safe \gls{ai} deployment through automated testing of new policies on real hardware before production deployment, rollback mechanisms for failed control actions, compare baseline versus learned policies, and constraint enforcement at the xApp level to prevent unsafe actions even from misbehaving \gls{ml} models.
    
\end{itemize}

\subsection{Future Research Directions}
Building on the foundation established in this work, we identify several promising directions for future research:
\begin{itemize}
    \item Predictive and Proactive Control: Leveraging \gls{lstm}, Transformer, or temporal convolutional networks to predict future network conditions (e.g., throughput degradation, congestion onset) and trigger preemptive control actions. For instance, extending the mobility scenario with predictive models that anticipate handover requirements before signal quality degrades, reducing connection interruptions.
    \item Multi-Agent Coordination: Deploying multiple collaborative xApps that coordinate control decisions across distributed \gls{ran} sites using multi-agent techniques.
    \item Transfer Learning and Sim-to-Real: Investigating transfer learning approaches where policies are pre-trained in simulation environments (e.g., ns-O-RAN \cite{lacava2023ns}) and fine-tuned on real deployments using \xdev testbed infrastructure.
    \item Integration with \gls{osc} \gls{ai}/\gls{ml} Framework: Developing tight integration between \xdev and the \gls{osc} \gls{ai}/\gls{ml} framework~\cite{aiframework-osc} to enable end-to-end model lifecycle management. This would support model training at the \nonrt \gls{ric}, distribution via the A1 interface, and deployment as \xdev-based xApps for near-real-time inference and control at the \nearrt \gls{ric}.
    \item Expanding Service Model Support: Extending \xdev to support additional E2 service models as they mature in the O-\gls{ran} specifications, such as \gls{ccc} for dynamic RAN parameter tuning and \gls{llc} for PHY-layer optimization,  providing even richer action spaces for \gls{ai} algorithms.
\end{itemize}

\subsection{Closing Remarks}

By bridging the gap between O-RAN specifications and practical xApp development, \xdev complements existing initiatives, such as the \gls{osc} \gls{ai}/\gls{ml} framework, while offering a lightweight, developer-oriented tool specifically focused on xApps. The framework's validation across heterogeneous stacks, integration with \gls{ci}/\gls{cd} workflows, and explicit design for \gls{ai}/\gls{ml} integration position it as a foundational infrastructure for the next generation of intelligent RAN control.

As O-RAN continues to evolve toward increasingly intelligent, software-driven networks, frameworks like \xdev play a crucial role in accelerating innovation, experimentation, and the realization of interoperable, closed-loop \gls{ran} control.

\bibliographystyle{IEEEtran}
\bibliography{references}
\balance
\end{document}